\documentclass[11pt]{article}

\usepackage{latexsym,amsthm,amsmath,amssymb,url}
\usepackage{subcaption}
\usepackage[inline]{enumitem}
\setlist[enumerate,1]{label=(\roman*),ref=(\roman*)}
\usepackage{tikz}
\usepackage{booktabs}
\usepackage{multirow}
\usepackage[authoryear,round]{natbib}

\newcommand{\Nmax}{N_{\max}}
\newcommand{\ymax}{\mathrm{Peak}}

\newcommand{\abs}[1]{\lvert#1\rvert}
\newcommand{\Integers}{\mathbb{Z}}

\newcommand{\set}[1]{\left\{#1\right\}} 

\usepackage[citecolor=blue, colorlinks]{hyperref}
\usepackage{doi}
\usepackage{graphicx}

\begin{document}

\title{The Effect of Social Distancing on the Reach of an Epidemic in Social Networks\protect\footnote{We would like to thank Matt Jackson, Emily Sharratt, and Joel Sobel for helpful comments. Two anonymous referees and an associate editor made many good suggestions that improved the paper considerably. Remaining errors are ours. The research of S.-H. H. was supported by the Ministry of Education of the Republic of Korea and the National Research Foundation of Korea (NRF-2019S1A5A8035341).}}
\author{
Gregory Gutin\protect\footnote{Computer Science Department, Royal Holloway, University of London.} \and Tomohiro Hirano\protect\footnote{Economics Department, Royal Holloway, University of London.} \and Sung-Ha Hwang\protect\footnote{College of Business, Korea Advanced Institute of Science and Technology (KAIST).} \and Philip R.\ Neary\protect\footnote{Economics Department, Royal Holloway, University of London.} \and Alexis Akira Toda\protect\footnote{Department of Economics, University of California San Diego.}
}

\date{\today}
	
\maketitle


\pagenumbering{arabic}


\begin{abstract}
\noindent
How does social distancing affect the reach of an epidemic in social networks? We present Monte Carlo simulation results of a \emph{Susceptible-Infected-Removed with Social Distancing} (SIRwSD) model. The key feature of the model is that individuals are limited in the number of acquaintances that they can interact with, thereby constraining disease transmission to an \emph{infectious subnetwork} of the original social network. While increased social distancing typically reduces the spread of an infectious disease, the magnitude varies greatly depending on the topology of the network indicating the need for policies that are network-dependent. Our results also reveal the importance of coordinating policies at the `global' level. In particular, the public health benefits from social distancing to a group (e.g., a country) may be completely undone if that group maintains connections with outside groups that are not following suit. 
\end{abstract}

\vspace{.1in}

\begin{tabular}{cl}
\emph{Keywords:} & SIRwSD model; Social Distancing; {infectious subnetwork};\\
& WS small world networks; BA scale-free networks.
\end{tabular}


\newpage
\section{Introduction}


How does social distancing affect the reach of an epidemic such as COVID-19 (Coronavirus Disease 2019) in social networks? To address this question we consider the idealised problem of the Susceptible-Infected-Removed (SIR) epidemic model \citep{KermackMcKendrick:1927:PRSLSCPMPC} in the presence of temporary social distancing constraints placed on the members of a society.


Our \emph{Susceptible-Infected-Removed with Social Distancing} model (SIRwSD model) is easily understood. As is standard, the vehicle that an infectious disease uses to spread is a \emph{contact network}:\ a graph, $G$, where vertices represent people and an edge between two vertices $i$ and $j$ captures the idea that person $i$ and person $j$ came into contact in such a way that the disease might pass between them.\footnote{The modelling epidemics on networks is well-established. See for example \cite{Pastor-SatorrasVespignani:2001:PRL} and \cite{MorenoPastor-Satorras:2002:EPJCMCS}. \cite{KissMiller:2017:} is a recent textbook treatment.} A social distancing policy is described by a function $\kappa$, defined on the set of vertices of $G$, that constrains the number of neighbours that each person may come into contact with.\footnote{Note that by `social distancing' we mean limiting the number of social interactions in which an individual partakes. We do not mean individuals maintaining a predetermined physical distance throughout a social interaction as is sometimes intended. Our measure is adapted from the game-theoretic model of Netflix Games \citep{GerkeGutin:2019:A,GutinNeary:2020:A}. There, the purchaser of a product may only share with a limited number of friends. Here, the carrier of an infectious pathogen is limited in the number of friends that they may interact with.} At any moment in time, the contact network $G$ and social distancing policy $\kappa$ together generate an \emph{infectious subnetwork}. Since the social distancing policy $\kappa$ temporarily deletes a subset of edges from $G$, there are simply less avenues along which the disease may be transmitted.



We run Monte Carlo simulations comparing the reach of the disease when there is no social distancing with that when social distancing measures are imposed.\footnote{All replication files are available at \href{https://github.com/alexisakira/COVID-19_network}{https://github.com/alexisakira/COVID\-19\_network}.} These comparisons are performed on three types of well-studied networks:\ the `random graphs' model of \cite{ErdosRenyi:1959:PMD} and \cite{Gilbert:1959:AMS} (hereafter, ERG), the `small world' networks of \cite{WattsStrogatz:1998:N} (hereafter, WS), and the `scale-free' model of \cite{BarabasiAlbert:1999:S} (hereafter, BA).\footnote{\label{fn:like-for-like}To ensure we are comparing like-for-like, we fix it such that all networks are connected and possess the same average degree. This means that it is the way in which the societies are organised, and not the overall level of interaction, that is being varied. 
}

We begin by focusing on two main questions for the constrained and unconstrained case:
\begin{enumerate*}
\item\label{item:q1} What is the likelihood that an infectious disease will become endemic?
\item\label{item:q2} What is the distribution of the peak infection rate over the lifecycle of the epidemic?
\end{enumerate*}
The reason for focusing on \ref{item:q1} is that the level of herd immunity attained is an important policy tool in knowing how, when, and by how much social distancing measures can be relaxed. The reason for focusing on \ref{item:q2} is that the peak infection rate corresponds to the most overloaded instance that a healthcare service encounters over the lifecycle of an epidemic. Success in both the above dimensions is not simultaneously possible because as one goes up the other goes down. Our results help to guide what the relative trade-offs are.

While the spread of an epidemic is curtailed when all individuals in a society face the same constraints, the reduction varies greatly depending on the topology of the social network. For all three social network structures, ERG, WS, and BA, that we consider, strong measures of social distancing (limiting everyone's daily interactions to 3 or fewer) stops an epidemic with high probability (see Figure \ref{fig:infection_network}). However, for societies structured according to WS, the fraction of individuals who are in state $R$ (`removed') after the epidemic has passed is much less over this range (see rows labeled `Herd Immunity' in Table \ref{t:large_sim}). Moderate social distancing (defined as 4-5 interactions per period) delays the peak of an epidemic but has little effect on the size of the peak and on the number that the disease ultimately reaches. The effect of mild social distancing (limiting individuals to 10 social interactions) differs greatly across network structures. With ERG networks, the effect is negligible. With WS networks, the outcome is the same as no intervention. With BA networks, the peak is significantly reduced. 


While we begin with comparing the outcomes of no social distancing with those to social distancing, our framework is flexible enough to address a host of other policy experiments. For example, we consider a network comprised of two densely connected components, interpreted as `countries', that have a small number of connections between them that we interpret as international friendships. We show that the public health benefits to a country that imposes strong social distancing measures are dramatically reduced, and perhaps eradicated entirely, if that country continues to allow international connections, interpreted as maintaining open borders, with a country that is not implementing similar measures. In a hyper-connected world this points to the need for `global' cooperation to eradicate an epidemic. In particular, if the global approach is uncoordinated with each country unilaterally applying social distancing measures without taking into account the policy choices of its neighbours, then an infectious disease may cycle around for far longer than otherwise desired. 

We also consider what happens when a subset of individuals are deemed `essential workers' who can go about their lives facing weaker social distancing constraints than the rest of the population. Our results show that even if only a small fraction of the population is deemed essential, the reach of the epidemic is similar to that wherein there are no essential workers. Finally, we consider another policy tool: time. Specifically, we consider a policy that begins with severe social distancing measures that are incrementally relaxed over time, and compare the outcome with that from a policy of mild but constant social distancing over a shorter window. We find that the public health outcome is better under gradual relaxation for ERG networks and WS networks, but worse for BA networks.


We conclude the paper by discussing how it fits in the literature and by suggesting some extensions to our framework that can be implemented in future work. However, we emphasise that the framework of this paper considers the benefits to social distancing when it is the only policy tool available. Clearly this is not realistic. In tackling real world epidemics policy makers have an array of tools available. For practical purposes governments need to understand how the many different epidemic management policies complement each other. This paper is about one such policy tool in isolation, but given its tractability we hope that it can be incorporated into richer models.

\section{The Model}\label{sec:Model}
%



Modelling how an infectious disease might spread through a population is done using a \emph{contact network}:\ an undirected graph, $G = (V, E)$, where $V$ is the set of vertices, and $E$ is the set of edges. Vertices represent individuals and an edge between two distinct vertices $i$ and $j$ captures the idea that these two people are acquaintances and meet in such a way that the disease may be transmitted from one to the other.



We assume throughout that $G$ is connected.  The neighbourhood of vertex $i$ in $G$ is denoted $N(i)$ and its degree is denoted $d(i) = \abs{N(i)}$. The \emph{capacity} of $G$ is a function $\kappa : V \to \Integers_{\geq 0}$. This capacity function $\kappa$ is our measure of social distance. A contact network $G$ and social distancing policy $\kappa$ together generate an \emph{infectious subnetwork}, $G^{\kappa}$. Intuitively, the policy $\kappa$ restricts the number of neighbours that each vertex can interact with, which will in turn cap the number of neighbours to whom any infected person can transmit the disease in a given period.\footnote{The mathematical machinery underpinning our model is adapted from the game-theoretic model of Netflix Games introduced in \citep{GerkeGutin:2019:A,GutinNeary:2020:A}. In a Netflix Game, each individual can be a Driver, $D$, who purchases netflix, or a Passenger, $P$, who free-rides on the purchase of neighbours, and the capacity function determines how many neighbours a Driver may share with. The Nash equilibria of a Netflix Game are characterised by a novel kind of spanning subgraph referred to as $DP$-\emph{Nash subgraph}. In the current environment, the capacity function dictates how many neighbours each individual may visit on a daily basis when a social distancing measures is in place.}




Our \emph{SIRwSD} {model} then operates as follows. Everyone in the population is currently in one of three states:\ \emph{Susceptible} ($S$) - has not had the disease and is therefore at risk; \emph{Infectious} ($I$) - currently has the disease and may therefore pass the disease to others; \emph{Removed} ($R$) - has had the disease and is no longer infectious (may be immune, isolated, dead, etc.). Time is discrete, starts at $t = 0$, and goes forever. Let $S_{it}$, $I_{it}$, $R_{it} \in \set{0, 1}$ be the status of individual $i$ at time $t$, where 1 means being that status and 0 means not being that status. (Clearly,  $S_{it} + I_{it} + R_{it} = 1$ for all $i \in V$ and all times $t$.) In every period, each infected vertex $i$ randomly selects $\kappa(i)$ of its neighbours, and, if any of the selected neighbours are in state $S$, then they become infected with probability $\beta \in (0, 1)$.\footnote{\label{fn:kappaNote}A few things to note. First, technically we should write that vertex $i$ selects $\min\set{\kappa(i), d(i)}$ neighbours since a vertex cannot select more neighbours than it has. Second, note that the model reduces to the standard networked SIR model when $\kappa(i) = d(i)$ for all $i \in V$. Third, when $0\leq \kappa(i) < d(i)$ for some $i \in V$, there are $\binom{d(i)}{\kappa(i)}$ ways for $i$ to select a subset of neighbours of size $\kappa(i)$. We assume a uniform distribution over the likelihood of choosing each subset of neighbours which implies no individual has a ``favoured neighbours''. This can of course be relaxed.} The probability of removal (moving from state $I$ to state $R$) is denoted by $\gamma \in (0, 1)$. We assume that $\gamma$ is constant over time and is the same for everyone. Once an individual enters state $R$, he stays there forever more. We say that the system has \emph{stopped} when there is no individual in state $I$.\footnote{There are continuous time versions of the SIR model. In such environments, the parameters $\beta$ and $\gamma$ are interpreted as rates or flows and are not probabilities constrained to lie in $(0, 1)$. Of course, when running simulations, one must discretise these continuous time models.}$^{,}$\footnote{The majority of theoretical papers on disease spread on networks focus on scenarios in which the disease will become endemic with probability 1 or will die out with probability 1. \cite{LancicAntulov-Fantulin:2011:PSMA} categorise where in $(\beta, \gamma)$ space both these events occur with positive probability.}


To illustrate how social distancing works in practice, we refer to Figure \ref{fig:socialDistance} below. This figure can be thought of ``zooming in'' on the local neighbourhood of individual $\ell$ who has 4 neighbours in the contact network. The leftmost image, shows precisely this. The next three images illustrate what transpires when a social distancing policy with $\kappa(\ell) = 2$ is imposed for three periods, with period $t=1$ being the first period in which the policy takes effect. In each period $t = 1, 2, 3$, individual $\ell$ randomly selects two of their four neighbours to interact with. The interactions that occur are depicted by solid lines while those that don't occur are depicted by dashed lines. For the realisation depicted, $\ell$ interacts with neighbours $h$ and $k$ in period 1, with neighbours $i$ and $k$ in period 2, and with neighbours $h$ and $i$ in period 3. Note that $\ell$ never encounters neighbour $j$ under this realisation. Thus the disease being transmitted (directly) from $\ell$ to $j$ is impossible while social distancing is in effect but would be possible without.

\begin{figure}[!htb]
\centering
\includegraphics[]{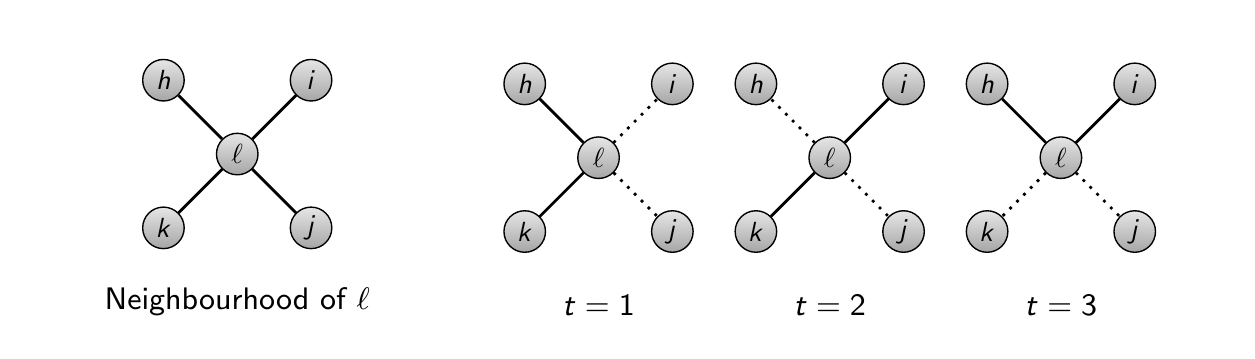}
\caption{The social distancing policy $\Nmax = 2$ in effect for three periods.}\label{fig:socialDistance}
\end{figure}



It is now straightforward to visualise how the SIRwSD model operates by extrapolating from what transpired in the neighbourhood of $\ell$ in Figure \ref{fig:socialDistance} to the entire contact network $G$. Assume the social distancing policy is given by $\kappa$, that it starts in period 1 and is in effect for $T$ periods. Denote the infectious subnetwork in period $t \in \set{1, \dots, T}$  by $G^{\kappa}_{t}$. For the duration of the policy, the capacity constrained SIR model is simply an SIR model operating on the sequence of infectious subnetworks $\set{G^{\kappa}_{1}, G^{\kappa}_{2}, \dots, G^{\kappa}_{T}}$.\footnote{Here we have assumed that the policy is constant over the window $\set{1, \dots, T}$. If the policy is allowed to vary with time, as in Subsection \ref{subsec:Relaxing}, we simply get a sequence of policies indexed by time, $\set{\kappa_{1}, \kappa_{2}, \dots, \kappa_{T}}$.}


This concludes the discussion of our SIR with Social Distancing model. Since the state of the system at any time $t$ is given by $\set{(S_{it}, I_{it}, R_{it})}_{i \in V}$, all that is required to run simulations are to specify the the environment, social distancing policy, and initial condition. This requires disease parameters $\beta$ and $\gamma$, a graph $G$, a sequence of capacity functions $\set{\kappa_{1}, \kappa_{2}, \dots, \kappa_{T}}$, and initial condition $\set{(S_{i0}, I_{i0}, R_{i0})}_{i \in V}$. With all this, the capacity constrained SIR model is a well-defined stochastic process that is easily simulated.





\section{Simulation Parameters}\label{sec:Parameters}

We run simulations on the capacity constrained SIR model described above for three different types of graphs:\ ERG, WS, and BA. We model one period as a day to calibrate network and epidemic parameters.

\subsection{Network Parameters}
We briefly review how each kind of graph is generated and the graph-parameters required to render them like-for-like. (See Footnote \ref{fn:like-for-like}.)

\paragraph{\cite{ErdosRenyi:1959:PMD} \& \cite{Gilbert:1959:AMS}}
These networks are often referred to simply as `random graphs'. Begin with the $n$ vertices and an empty edge set. Now, for every pair of distinct vertices $i$ and $j$, form edge $ij$ with fixed probability $p$. The expected number of edges in the resulting graph is $\frac{1}{2}n(n-1)p$, because there are $\binom{n}{2}=\frac{1}{2}n(n-1)$ potential edges and each edge is realised with probability $p$.

\paragraph{\cite{WattsStrogatz:1998:N}}
The resulting network is often referred to as a `small world'. Begin with the $n$ vertices  connected in a ring lattice where each vertex has $k$ neighbours to the left and $k$ neighbours to the right, where $k \ll n$. Proceed clockwise around the ring one time, and for each vertex, rewire every edge that it has with the $k$ vertices immediately to its right with rewiring probability $q$. Note that since the net change in the number of edges is zero, the number of edges in the resulting graph remains exactly $nk$, because each vertex has $2k$ neighbours and each edge is counted twice.

\paragraph{\cite{BarabasiAlbert:1999:S}}
The resulting network is often referred to as `scale free'. Begin with a complete graph on $m_0 \ll n$ vertices and allow time to increment forward from $t=1$ to $t = n-m_{0}$. At each point in time, a new vertex is born and the newly born vertex forms one edge with $m \leq m_{0}$ of the existing vertices, where the probability that the newly born vertex connects to existing vertex $i$ is given by $d(i)/\sum_{j}d(j)$. In total  $n-m_{0}$ new vertices are added so the number of edges in the resulting graph is always exactly $\frac{1}{2}m_0(m_0-1)+m(n-m_0)$.\\


We choose to equalise average degree, $\bar{d}$, across all network types. This ensures we are comparing like-for-like since it is the way in which the societies are organised, and not the overall level of interaction, that is being varied. Given that the number of edges in each graph will be $n\bar{d}/2$, some straightforward algebra yields the following parameter requirements: $p = \frac{\bar{d}}{n-1}$ for ERG, $k=\frac{\bar{d}}{2}$ for WS, and $m=\frac{n\bar{d}-m_0(m_0-1)}{2(n-m_0)}$ for BA. 

In each network we fix the number of vertices to be $n =1{,}000$ and set $\bar{d} = 10$.\footnote{\cite{ZhaoyangSliwinski:2018:PA} find that the average number of daily interactions for adults is 12 although it varies with age. Our choice of $\bar{d} = 10$ is made so as to be in line with this.} Such a choice requires setting the ERG parameter $p = 0.01$. For WS we choose rewiring probability $q = 0.1$, which is relatively standard, and $k = \bar{d}/2 = 5$. For the BA network, for simplicity we set $m_{0} = 1$, and round up $m$ so that $m=\bar{d}/2=5$.

\subsection{Epidemic Parameters}

In the absence of a cure, the removal probability $\gamma$ is a biological parameter determined by the infectious disease. Through contact tracing in Wuhan, China, \cite{Li_2020} estimate the mean serial interval for COVID-19, which corresponds to $1/\gamma$, to be 7.5 days. We round $\gamma$ to the first significant digit and set $\gamma=0.1$. To calibrate $\beta$, we proceed as follows. Since average degree is $\bar{d}$, an infected individual will infect $\beta\bar{d}$ others in one period and on average $\beta\bar{d}/\gamma$ over the infectious period when social distancing is unconstrained. The expression $\beta\bar{d}/\gamma$ must equal the basic reproduction number $\mathcal{R}_0$ of the infectious disease, which is estimated to be around 3 in \cite{Toda:2020:} for COVID-19.\footnote{These epidemic parameters may change with the accumulation of new scientific knowledge. Some of the parameters are updated at the Center of Disease Control and Prevention (CDC) website at \url{https://www.cdc.gov/coronavirus/2019-ncov/hcp/planning-scenarios.html}. As of September 2020, the best estimate of $\mathcal{R}_0$ is 2.5, with a possible range of 2.0--4.0.} Given our choice of $\bar{d}=10$ and $\gamma=0.1$, we have that $\beta = 0.03$. Finally, we initialise the system by setting each vertex to state $I$ with probability $y_{0} = 0.01$ (so that, on average, each trial starts with 1\% of the population infected).

To verify that our calibration of the transmission probability $\beta$ is correct, we simulate the effective reproduction number $\mathcal{R}_t$ in our social network model. In the classical SIR model of \cite{KermackMcKendrick:1927:PRSLSCPMPC} (in continuous-time and homogeneous interaction), the evolution of the fraction of infected agents $y_t$ satisfies the differential equation
\begin{equation}
\dot{y}_t=\beta x_ty_t -\gamma y_t=\gamma (\beta x_t/\gamma -1)y_t=\gamma (\mathcal{R}_t-1)y_t,\label{eq:Rt1}
\end{equation}
where $\beta>0$ is the transmission rate, $\gamma>0$ is the recovery rate, $x_t\in [0,1]$ is the fraction of susceptible agents, and $\mathcal{R}_t=\beta x_t/\gamma$ is called the effective reproduction number. Dividing both sides of \eqref{eq:Rt1} by $y_t>0$, after some algebra we obtain
\begin{equation}
\mathcal{R}_t=1+\frac{1}{\gamma}\frac{\mathrm{d}}{\mathrm{d}t}\log y_t.\label{eq:Rt2}
\end{equation}
Using the analogy from \eqref{eq:Rt2}, in our discrete-time model, we may \emph{define} the effective reproduction number by
\begin{equation}
\mathcal{R}_t=1+\frac{1}{\gamma}(\log y_t-\log y_{t-1})=1+\frac{1}{\gamma}\log \frac{y_t}{y_{t-1}}.\label{eq:Rt3}
\end{equation}

\begin{figure}[!htb]
\centering
\includegraphics[width=0.7\linewidth]{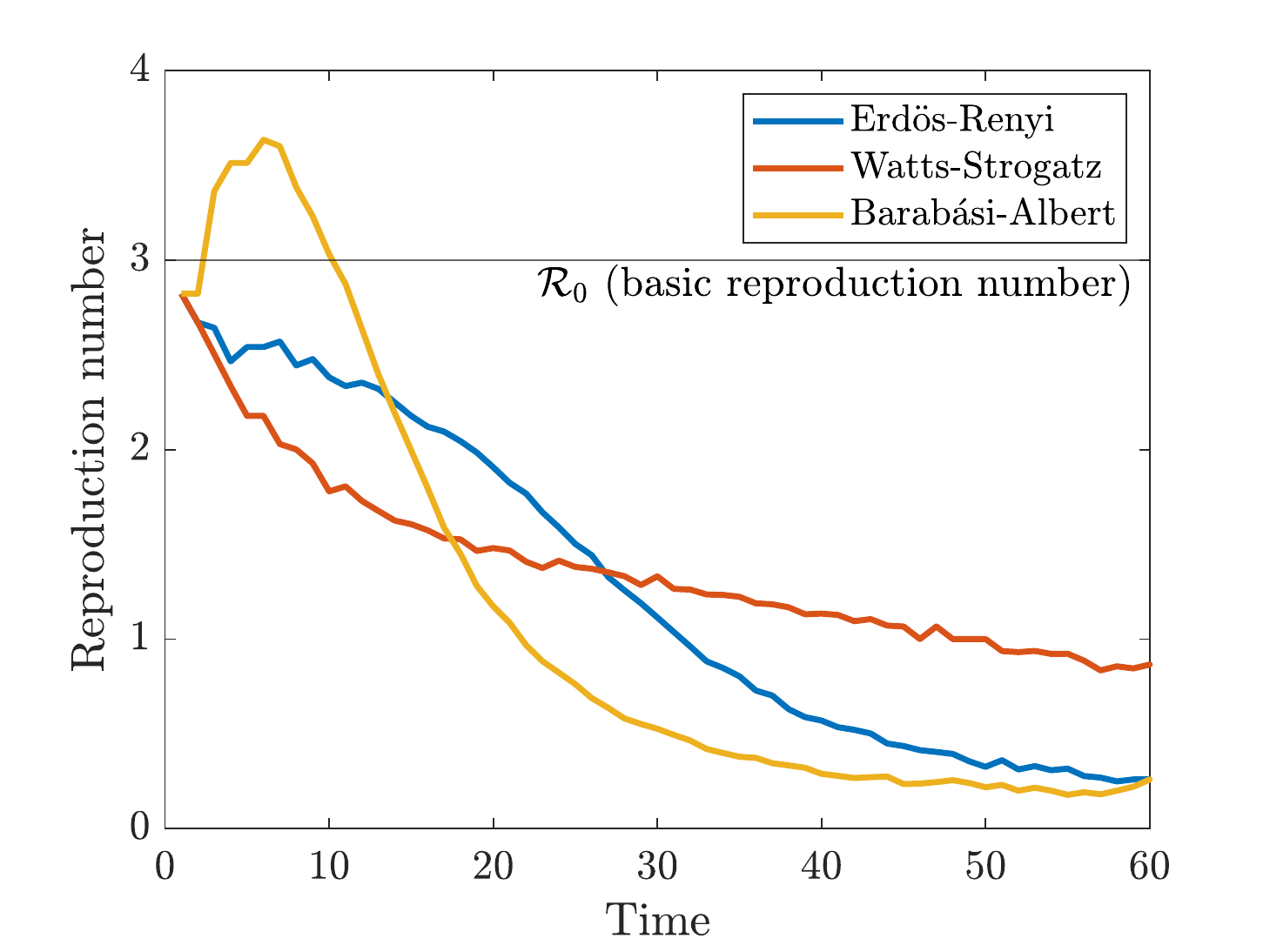}
\caption{Median effective reproduction number.}\label{fig:R0}
\end{figure}

Figure \ref{fig:R0} plots the median effective reproduction number \eqref{eq:Rt3} across 1,000 simulations described below for each network type. Consistent with our construction, the simulated effective reproduction number is close to the theoretical value $\mathcal{R}_0=3$ at the beginning of the simulation. Therefore we confirm that we are comparing networks like-for-like.\footnote{We thank an anonymous referee for bringing this important point to our attention.}

\section{Simulation Results}\label{sec:Results}

While a social distancing measure $\kappa$ may assign individual-specific constraints, in this section we will assume that it does not.\footnote{Many interesting policy experiments stipulate that different individuals face different social distancing constraints. We study one such scenario in Subsection \ref{subsec:Essential} and propose others in Section \ref{sec:Extensions}.} This allows us to abuse notation somewhat by simply writing $\Nmax$ for the maximum number of neighbours that each individual will interact with.\footnote{Caveat:\ as per Footnote \ref{fn:kappaNote}, when we say that ``all vertices face constraint $\Nmax$'', in actuality every vertex $i$ is assigned capacity $\kappa(i) = \min\set{\Nmax, d(i)}$.}

\begin{figure}[!htb]
\centering
\begin{subfigure}{0.48\linewidth}
\includegraphics[width=\linewidth]{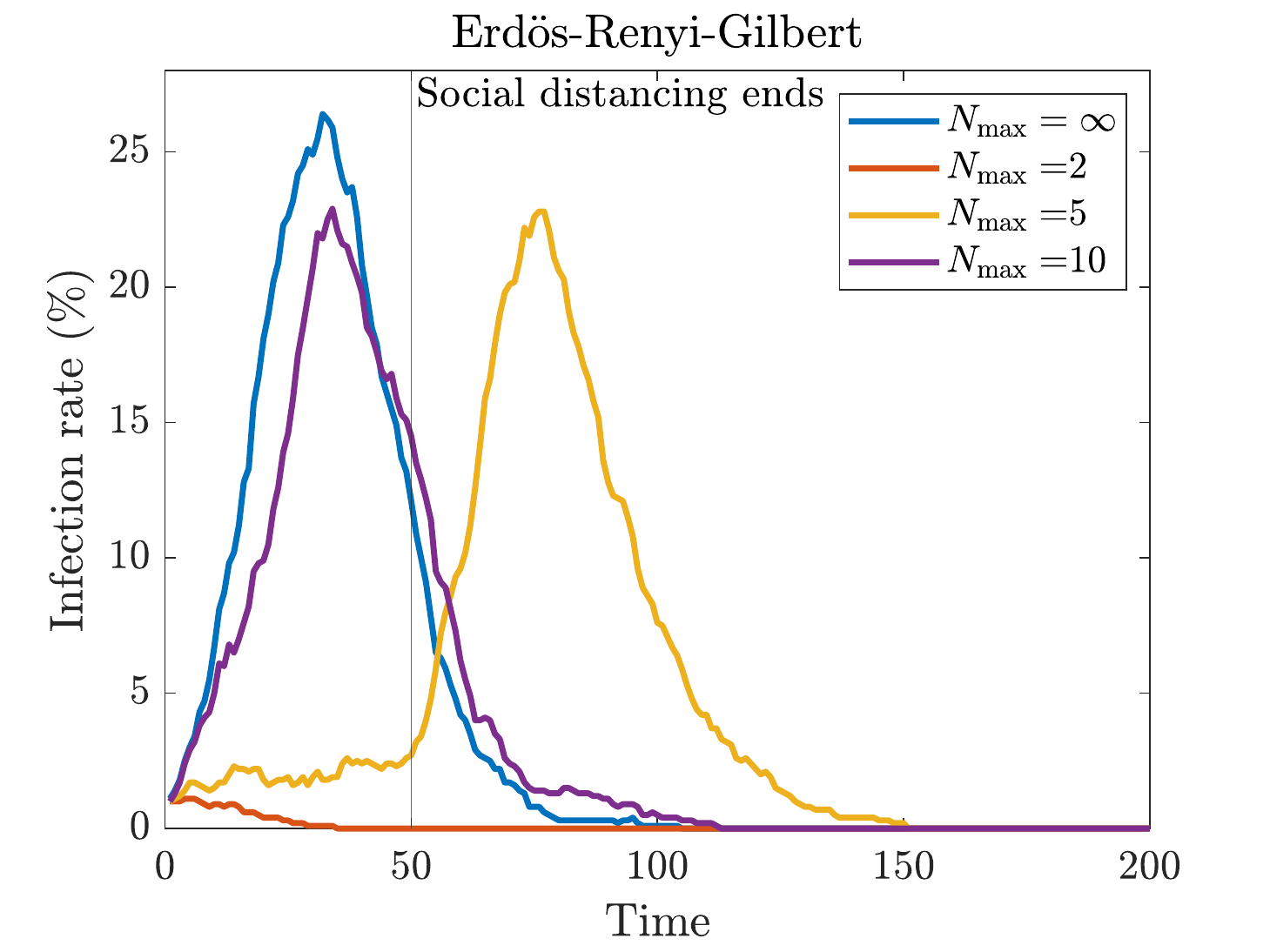}
\end{subfigure}
\begin{subfigure}{0.48\linewidth}
\includegraphics[width=\linewidth]{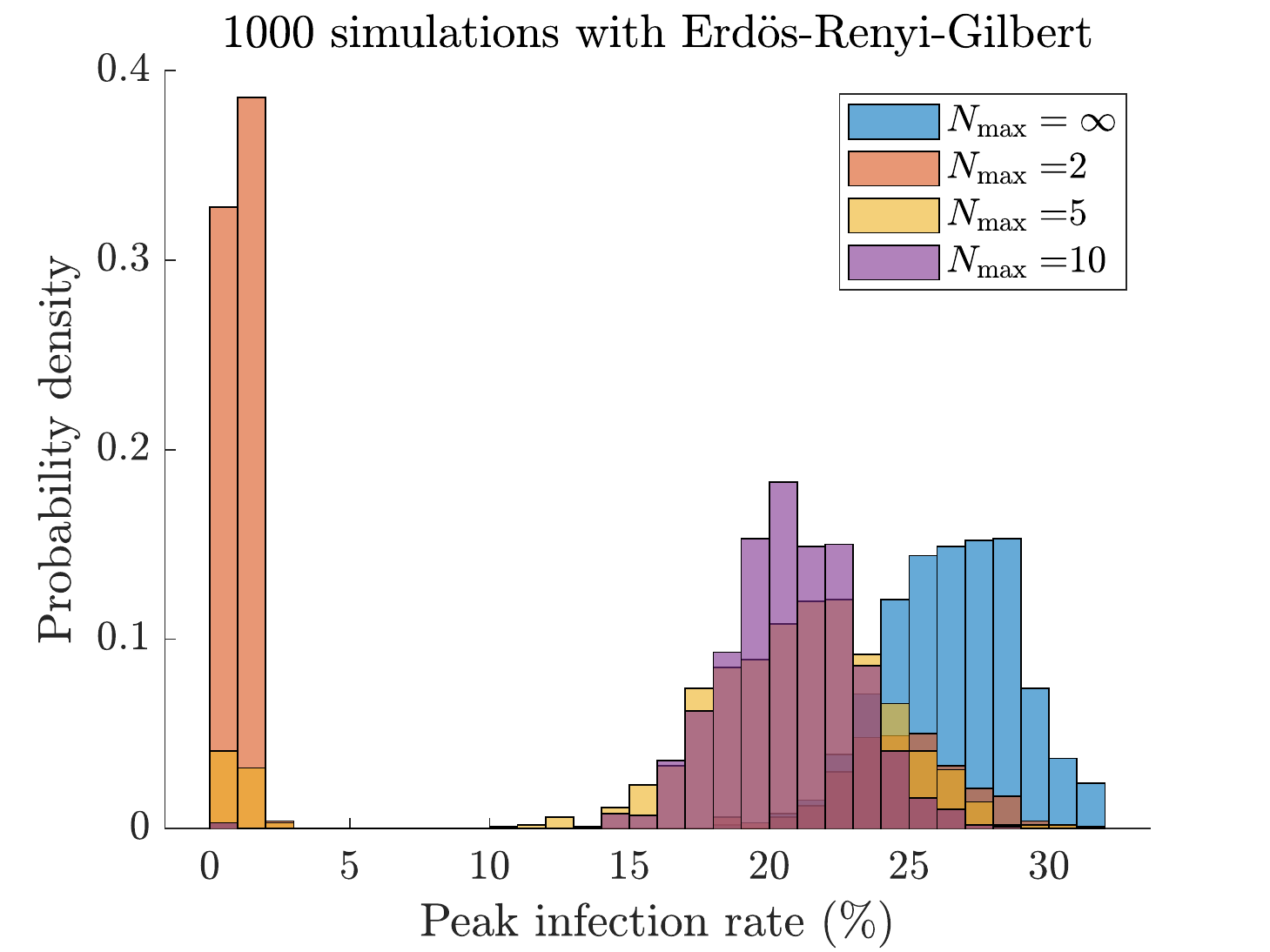}
\end{subfigure}
\begin{subfigure}{0.48\linewidth}
\includegraphics[width=\linewidth]{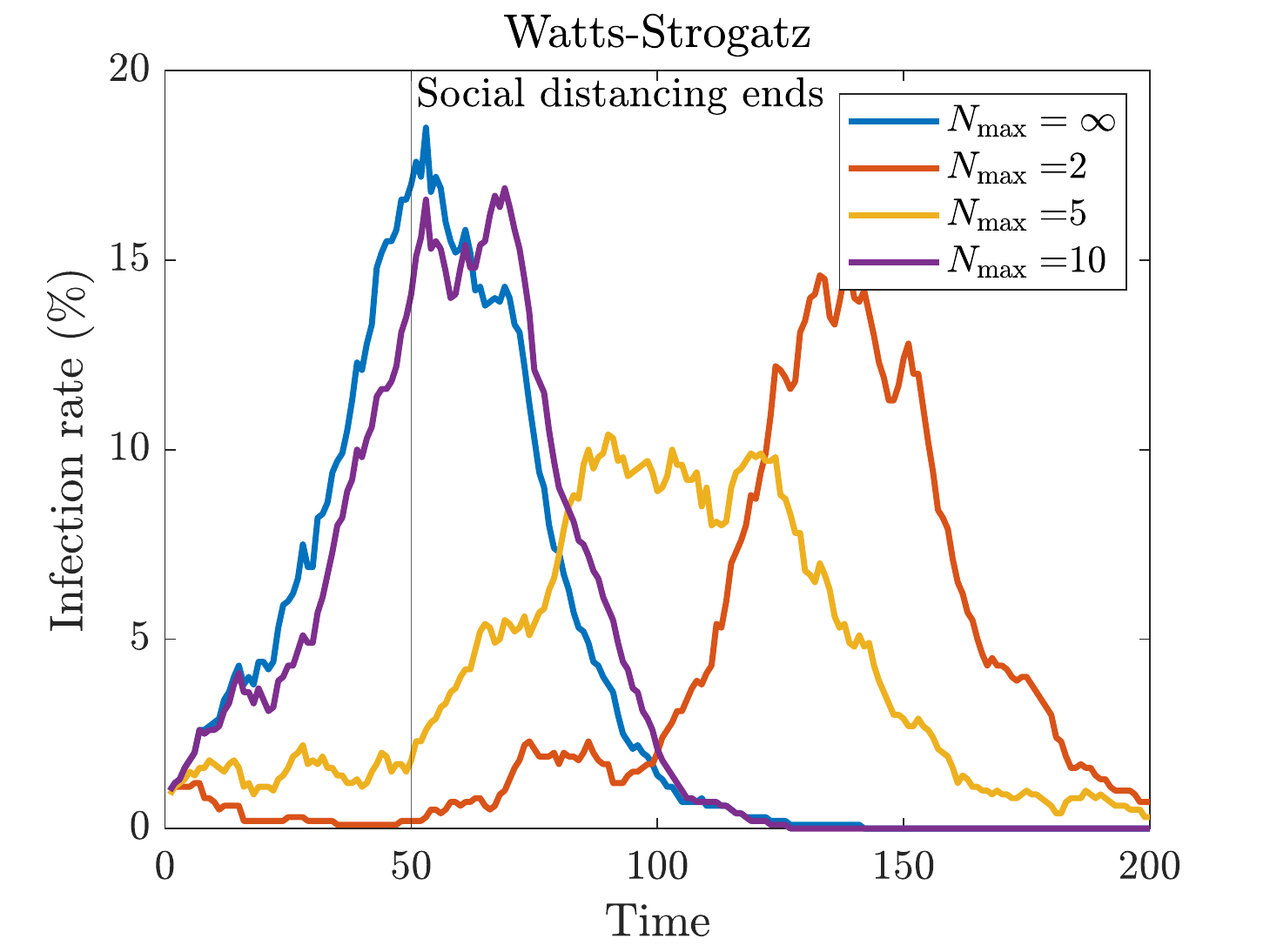}
\end{subfigure}
\begin{subfigure}{0.48\linewidth}
\includegraphics[width=\linewidth]{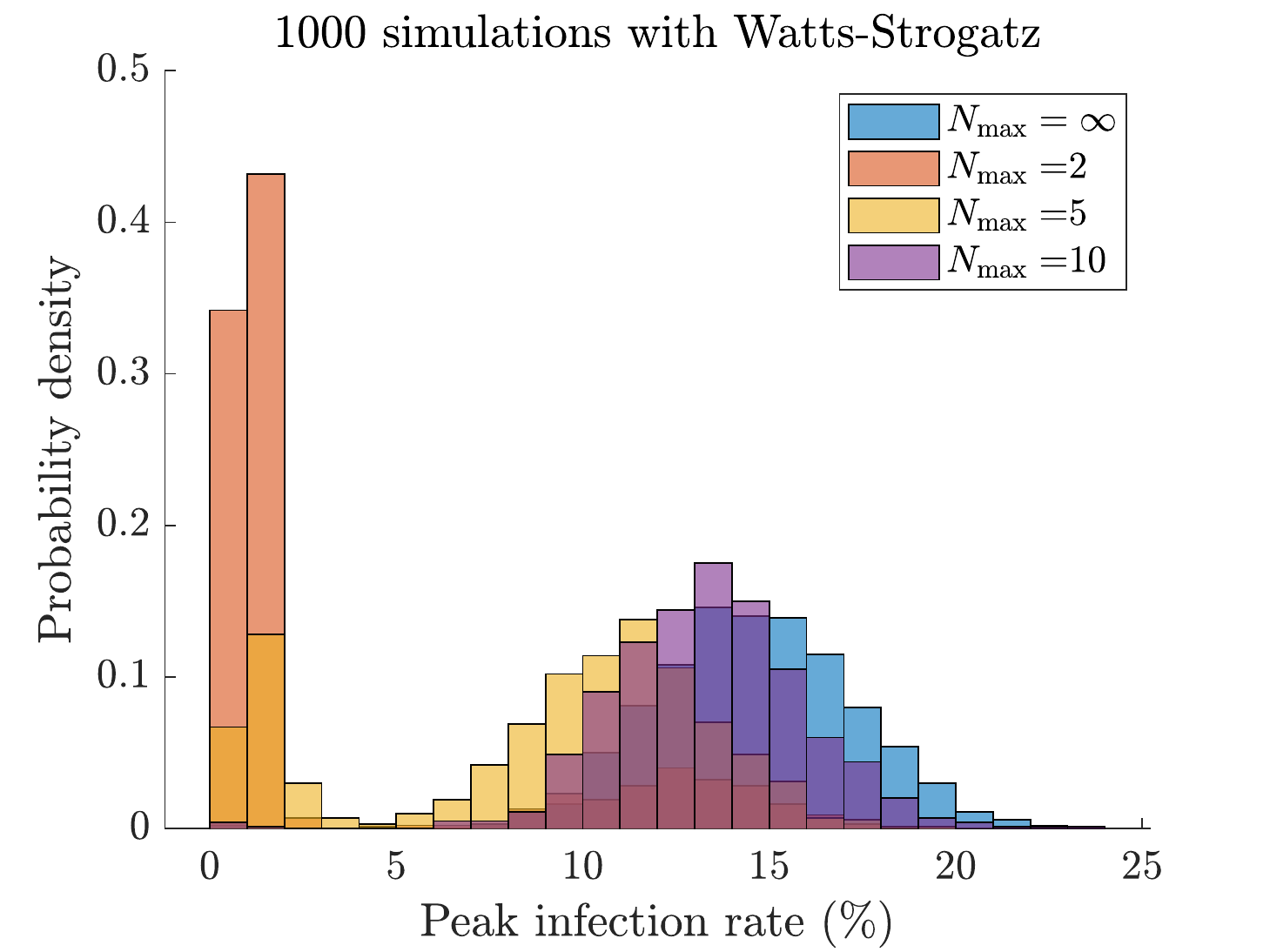}
\end{subfigure}
\begin{subfigure}{0.48\linewidth}
\includegraphics[width=\linewidth]{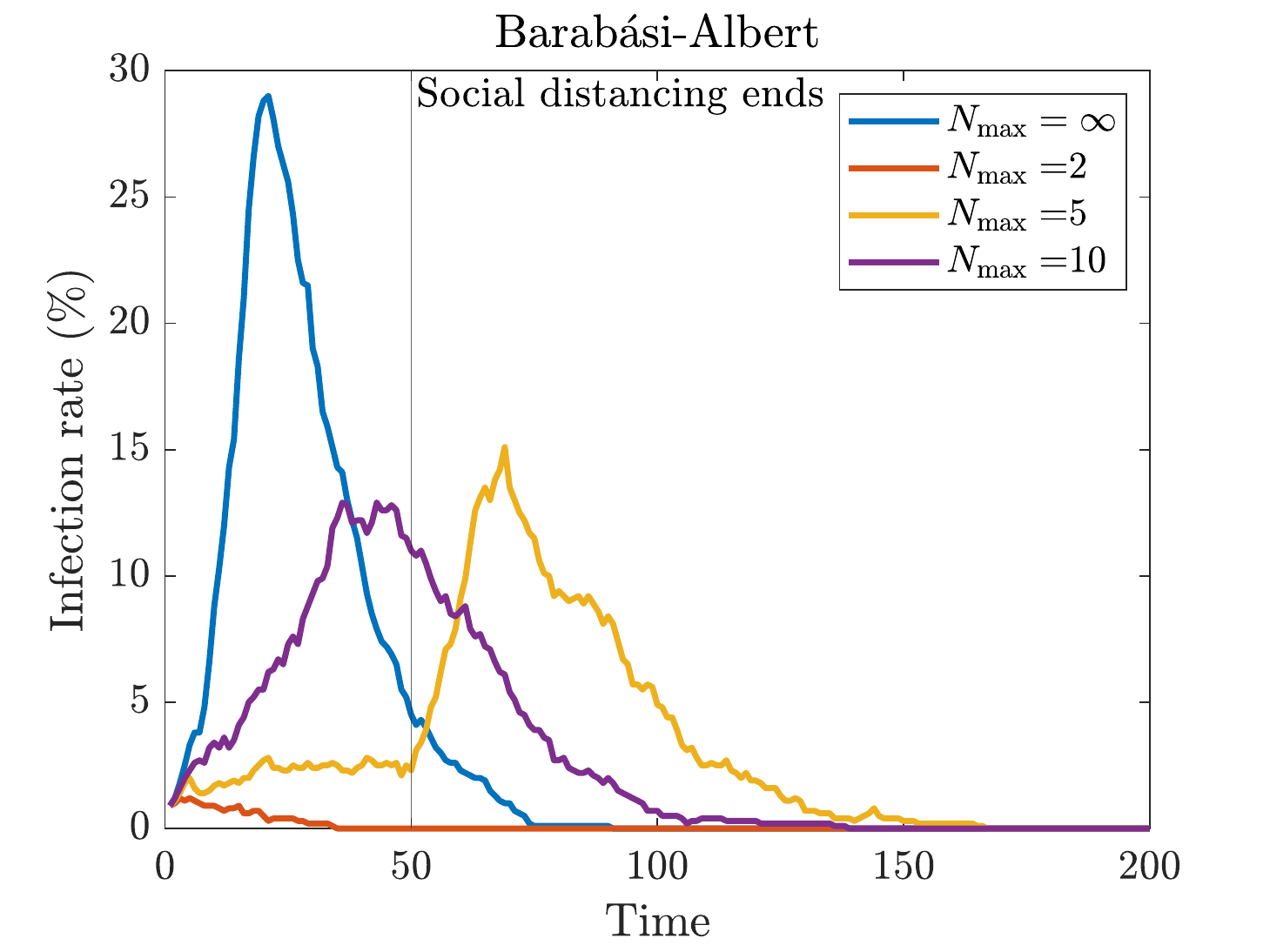}
\end{subfigure}
\begin{subfigure}{0.48\linewidth}
\includegraphics[width=\linewidth]{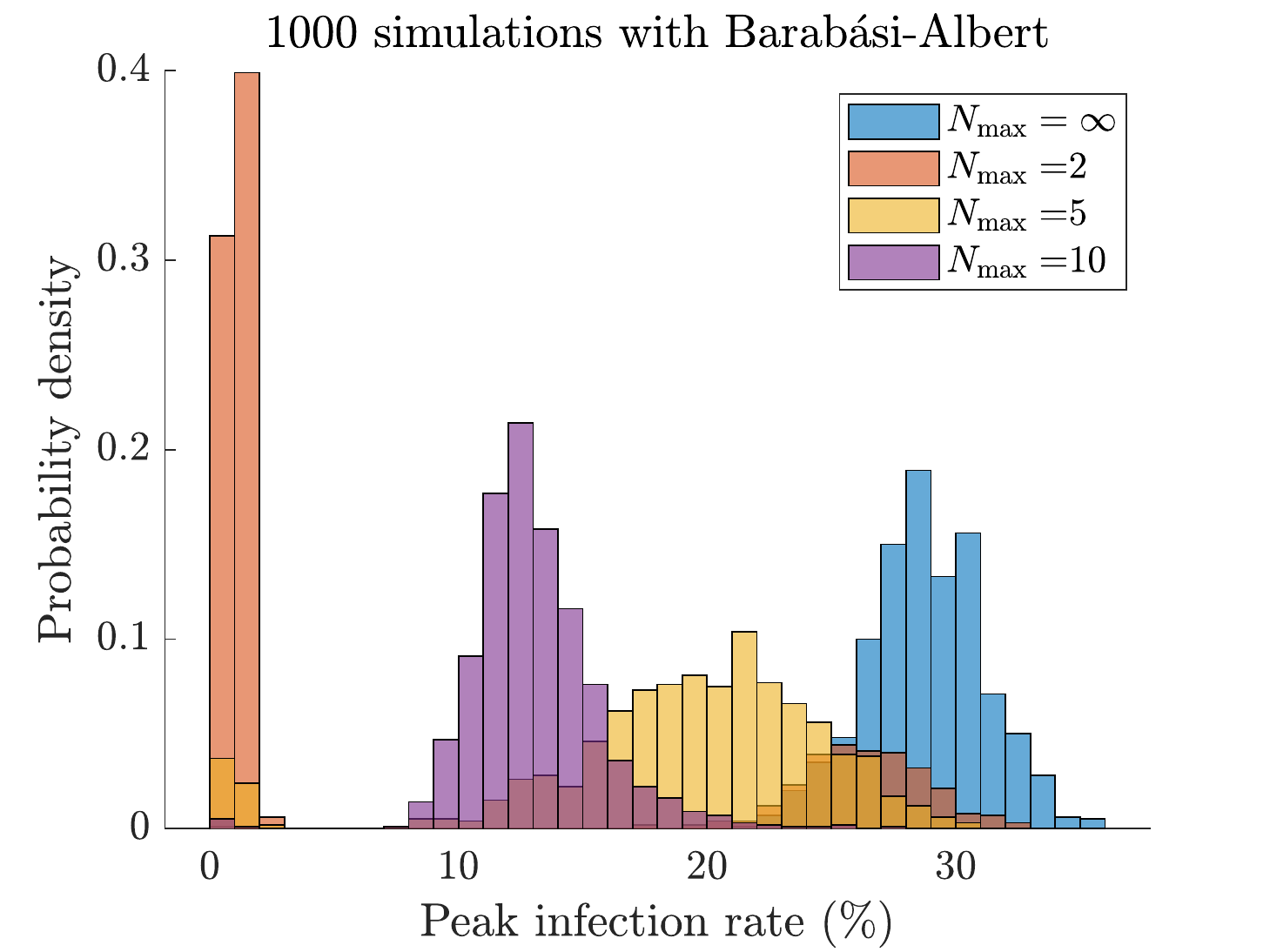}
\end{subfigure}
\caption{Results for one trial (left) and 1{,}000 trials (right).}\label{fig:infection_network}
\end{figure}

We allow the social distancing measure $\Nmax$ to take values in the set $\set{1, 2, 3, 4, 5, 10, \infty}$, where the value $\infty$ denotes the standard SIR model (i.e., our model without any constraints). Lastly, unless otherwise stated, we suppose that the social distancing measure starts in period $1$ and is lifted from period 51 onwards. Figure \ref{fig:infection_network} below contains six panels organised in a $3\times2$ format. The first row refers to ERG networks, the second to WS networks, and the last row to BA networks.

The left panel in each row presents results of four single trials where the trials are distinguished by the value of $\Nmax = 2, 5, 10$, and $\infty$ (we choose not to present all values of $\Nmax$ as the resulting image is too cluttered). The horizontal axis is time and the vertical axis is infection rate in the population. The vertical line at $t=50$ represents the lifting of the social distancing restriction.

The right panel in each row presents a histogram of peak infection rates computed from 1{,}000 simulations of the type shown in the left panel. To see the connection between the two panels, we note that each left panel provides one data point for the right panel. To further cement understanding, note that in the left panel of the WS row, the peak infection rate for $\Nmax = 2$ trial is greater than that for the $\Nmax = 5$ trial.

We begin with the left panels referring to individual trials. For the ERG and BA networks, the social distancing measure of $\Nmax = 2$ eradicates the epidemic within the 50 period window. Note however that the $\Nmax = 2$ trial does not eradicate the epidemic for the WS network. For each of the three network structures, the trial with moderate social distancing, $\Nmax=5$, delays the peak of the epidemic but has little impact on the size of the peak. The effect of mild social distancing, $\Nmax=10$, differs across network structures. With a random network (ERG), the effect is mild. With a small world network (WS), the outcome appears equivalent to no intervention ($\Nmax = \infty$). This is because in the WS network, each agent has $\bar{d}=10$ neighbours, so the restriction $\Nmax=10$ does not bind. (The histograms differ only due to sampling error.) With a power law network (BA), the peak is significantly reduced. This is because in the BA network there are a small number of individuals with very high degree. So, since many of the edges of the contact network have a high degree individual at one end, the same holds true for the infectious subnetwork regardless of what it may be. This means that these `hubs' are very likely to become infected even under social distancing. However, due to social distancing, an infected hub infects nowhere near as many as when there are no constraints. These hubs no longer act as ``super-spreaders''.

While the findings in the left panel of Figure \ref{fig:infection_network} are illustrative, they represent only a single trial and may not be representative. As such, to evaluate the robustness, we now turn our attention to the panels in the right hand column. Here each panel presents histograms of the peak infection rate attained in 1{,}000 trials for a given network type (four histograms on each graph - one for each of the same measures of social distancing as in the left panels).\footnote{We consider peak infection rate as it captures the worst case scenario. This has been cited as a particularly important measure in the COVID-19 epidemic where it is often the excess numbers of sick individuals that strain a healthcare system (limited numbers of intensive care beds, ventilators, and personal protective equipment for medical professionals) that can be as important an issue to public health as any other factor.} For all three network structures, setting $\Nmax=2$ eradicates the disease in $\sim 70\%$ of the trials. We note that the empirical frequency distribution of peak infection is bimodal for the value $\Nmax = 2$ with the lower mode corresponding to the epidemic being eradicated.  Setting $\Nmax =5$ eradicates the disease in slightly more than $\sim$$10\%$ of the trials of the WS networks and even less for ERG and BA. Setting $\Nmax = 10$ almost never eradicates the disease for any kind of network. Setting $\Nmax = \infty$ never eradicates the disease for any kind of network.

Conditional on the disease not being eradicated, the distribution of peak infection rate is the same as no intervention for ERG and WS. This is because the epidemic restarts after the lifting of social distancing measures. However, for the BA network, conditional on the disease not being eradicated the distribution of peak infection rate for $\Nmax = 5$ is sandwiched between that of $\Nmax = 10$ and $\Nmax=\infty$. At first glance this may appear mysterious but in fact it is not. The explanation is that the distribution of peak infection rate presented says nothing about the precise moment, during a particular trial, that the peak infection rate was attained. For $\Nmax = 5$, the peak infection rate will typically occur \emph{after} the social distancing restrictions have been lifted (as can be seen for the trial in the left panel of ERG), but this is not the case for $\Nmax = 10$. For BA networks, it would appear that setting $\Nmax = 10$ is a superior policy to the policy to setting $\Nmax = 5$.


Table \ref{t:large_sim} presents further results. The variable `$\ymax$' denotes the peak infection rate. {`Std.' and `Med.' denote its standard deviation and median, respectively. `Removed' denotes the fraction of population that were removed (has recovered) by the end of the epidemic, which also equals the cumulative number of infections.} Letting $T$ denote the last period before the system stops, we say that the society has acquired \emph{herd immunity} if $\beta \bar{d} x_T\le \gamma$, where $x_T$ is the fraction of susceptible individuals at time $T$. {This condition roughly says that a new infection no longer causes an exponential growth in cases.} (See \cite{Toda:2020:} for a discussion.) {The variable `Herd immunity' denotes the fraction of simulations in which herd immunity thus defined has been acquired.}



{As can be seen from Table \ref{t:large_sim}, there is a trade-off to be made between acquiring herd immunity and keeping the peak low. In most specifications, the probability of acquiring herd immunity and the average peak are both increasing in $\Nmax$. This suggests that drastic social distancing policies are fragile in the sense that while they tend to contain the disease, it is less likely to achieve herd immunity and the society is susceptible to a recurrent epidemic. Furthermore, note that the variable `$\overline{\ymax}\mid \text{Immunity}$' (meaning `average peak conditional on herd immunity being attained') is decreasing in $\Nmax$. This shows that when the disease control has failed, the peak is actually worse (higher) under more strict social distancing policies.}

\begin{table}[!htb]
\centering
\caption{Effectiveness of non-pharmaceutical interventions in networks.}\label{t:large_sim}
\begin{tabular}{clrrrrrrr}
\toprule
Network & Variable & \multicolumn{7}{c}{$\Nmax$} \\
& & $\infty$ & 1 & 2 & 3 & 4 & 5 & 10\\
\cmidrule(lr){1-1}
\cmidrule(lr){2-2}
\cmidrule(lr){3-9}
\multirow{6}{*}{ERG} & $\overline{\ymax}$ & 26.5 &   3.4 &   7.7 &  13.9 &  18.7 &  19.5 &  20.7\\
& Std.($\ymax$) & 2.8&  7.4& 10.7& 11.4&  8.9&  6.1&  2.5\\
& Med.($\ymax$) & 26.6&  1.0&  1.2& 20.9& 22.0& 20.8& 20.7\\
& Removed & 87.1& 10.2& 25.8& 50.1& 70.1& 78.8& 83.5\\
& $\overline{\ymax}\mid \text{Immunity}$ & 26.6& 25.1& 24.7& 23.6& 22.8& 21.1& 20.7\\
& Herd immunity & 99.7& 10.3& 28.2& 56.9& 80.9& 92.4& 99.7\\
\cmidrule(lr){1-1}
\cmidrule(lr){2-2}
\cmidrule(lr){3-9}
\multirow{6}{*}{WS} & $\overline{\ymax}$ & 14.5& 1.9& 3.5& 5.1& 7.3& 8.8&13.3\\
& Std.($\ymax$) & 2.8& 3.2& 4.8& 5.4& 5.3& 4.6& 2.6\\
& Med.($\ymax$) & 14.5& 1.0& 1.1& 1.5& 8.9&10.1&13.3\\
& Removed & 82.9& 7.5& 18.0& 29.7& 46.3& 59.8& 81.5\\
& $\overline{\ymax}\mid \text{Immunity}$ & 14.6&12.6&12.6&12.2&11.7&11.3&13.4\\
& Herd immunity & 99.4& 7.2&19.9&33.5&54.8&72.1&98.7\\
\cmidrule(lr){1-1}
\cmidrule(lr){2-2}
\cmidrule(lr){3-9}
\multirow{6}{*}{BA} & $\overline{\ymax}$ & 28.7&  4.0&  8.1& 13.8& 18.2& 18.9& 13.0\\
& Std.($\ymax$) & 2.5&  8.4& 11.4& 12.0&  9.3&  6.2&  2.6\\
& Med.($\ymax$) & 28.7&  1.0&  1.2& 20.3& 21.6& 19.9& 12.7\\
& Removed &80.7& 10.4& 23.6& 43.3& 60.9& 69.8& 66.1\\
& $\overline{\ymax}\mid \text{Immunity}$ & 28.7& 27.1& 26.2& 24.6& 22.8& 20.4& 14.1\\
& Herd immunity & 99.9& 11.5& 28.2& 54.1& 78.6& 89.7& 49.2\\
\bottomrule
\end{tabular}
\caption*{\footnotesize Note: ``$\overline{\ymax}$'' is the mean peak infection rate (\%). ``Std.''\ and ``Med.''\ are the standard deviation and median of peak infection rates across 1000 simulations. ``Removed'' is the fraction (\%) of population removed (recovered) by the end of the epidemic. ``$\overline{\ymax}\mid \text{Immunity}$'' is the mean peak infection rate conditional on acquiring herd immunity. ``Herd immunity'' is the fraction of simulations (\%) in which herd immunity was acquired.}
\end{table}

Perhaps the most surprising result in Table \ref{t:large_sim} is the value of the variable `Herd immunity' for BA networks when $\Nmax=10$. The value it takes is 49.2, which is considerably less than the value it takes when $\Nmax=5$. The reason for this is that when $\Nmax=10$, the potential `super spreaders' (those with very high degree) are more likely to become infected during the social distancing window than when $\Nmax=5$. But given the vertices with high degree become infected during the measures, they are likely to be recovered when the social distancing restriction are lifted. As such, these vertices do not maximise their reach.

In summary, choosing a small $\Nmax$ such as $\Nmax=1,2$ reduces the peak but prevents building herd immunity, which makes the society susceptible to further epidemics. Choosing an intermediate $\Nmax$ such as $\Nmax=3,4,5$ does not necessarily reduce the peak, while mildly preventing herd immunity. Choosing $\Nmax=10$ generally reduces the peak infection rate \emph{and} achieves herd immunity.

\section{Policy Issues}\label{sec:PolicyQs}

Our setup is sufficiently flexible that we can address a wide range of policy issues. In this section we discuss some extensions and present the results.

\subsection{Tackling Epidemics Requires `Global' Cooperation}\label{subsec:Global}

Suppose there are two countries, Country $A$ and Country $B$. Each country has a population of $n$ individuals, and when viewed in isolation each country forms its own connected network. Suppose further that every pair of individuals from different countries are randomly connected with fixed probability $1/(10n)$, and we interpret connections of this form as international relationships. (We choose a low value so that in expectation only a very small minority of each individual's interactions are with foreigners.)

For the simulations, we set $n=1{,}000$ and $\bar{d} = 10$ for each country, and we assume that both countries have the same social structure (both ERG, both WS, or both BA). We use the same values for $\beta$, $\gamma$, and initial infection probability $y_{0}$, as before.\footnote{We assume that in both countries 1\% of individuals are infected. Of course it is possible to address a host of other related questions. An example would be, ``Suppose $x$\% of individuals in Country $A$ are infected while nobody in Country $B$ is infected. What then is the lowest value of $x$ such that if $x$\% of individuals in Country $A$ are infected then the reach of the epidemic will be both Country $A$ and Country $B$?''.}  However, instead of assuming that temporary social distancing measures are imposed on everyone as in the simulations of Section \ref{sec:Results}, we now suppose that Country $A$ levies social distancing measures while Country $B$ imposes no measures of any kind. Figure \ref{fig:twocountry} below depicts the simulation results for one trial of the infection rate in both countries as a function of time for the range of different social distancing measures in Country $A$ given by $\Nmax = 2, 5, 10$, and $\infty$. 

\begin{figure}[!htb]
\centering
\begin{subfigure}{0.48\linewidth}
\includegraphics[width=\linewidth]{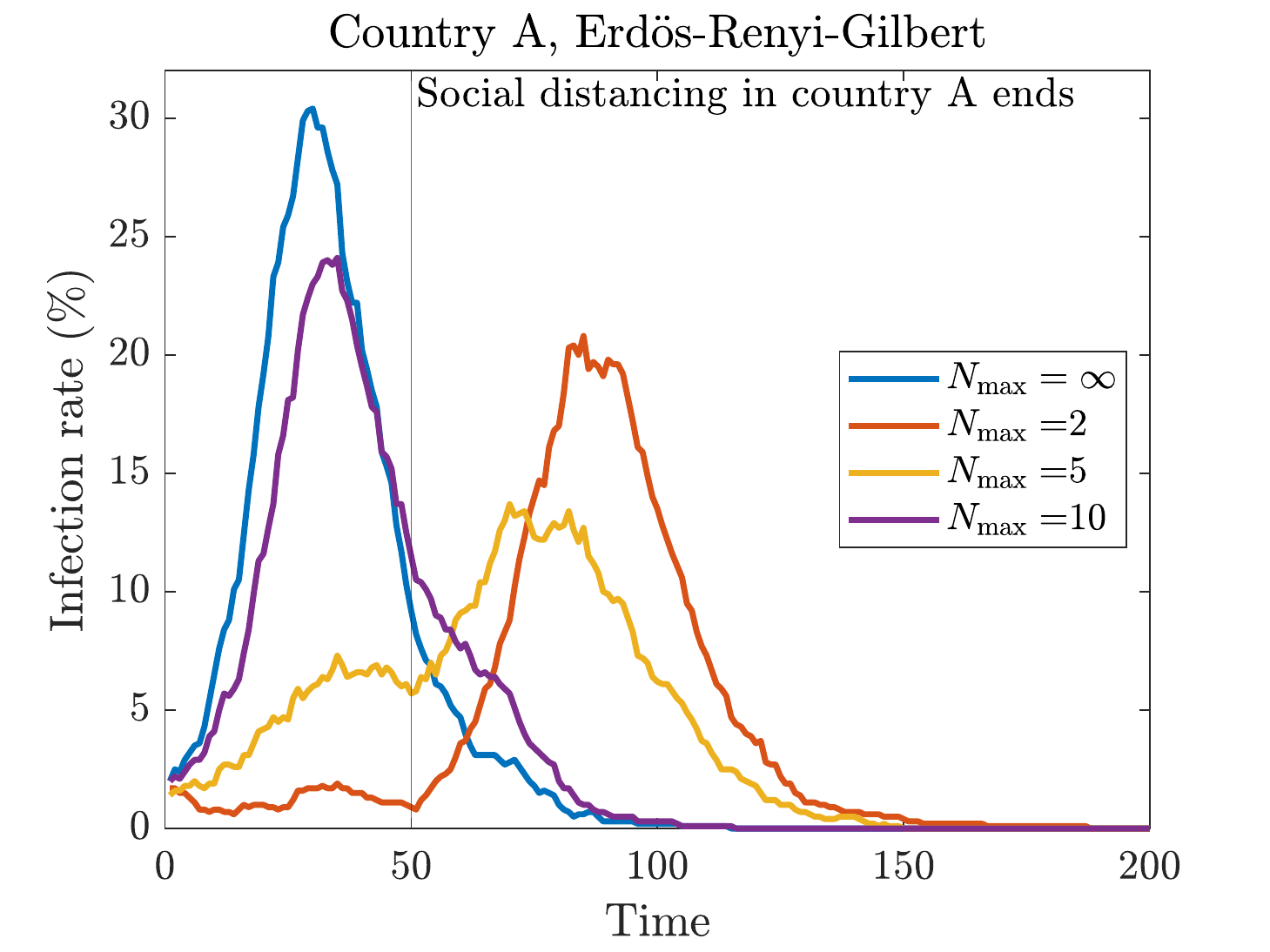}
\end{subfigure}
\begin{subfigure}{0.48\linewidth}
\includegraphics[width=\linewidth]{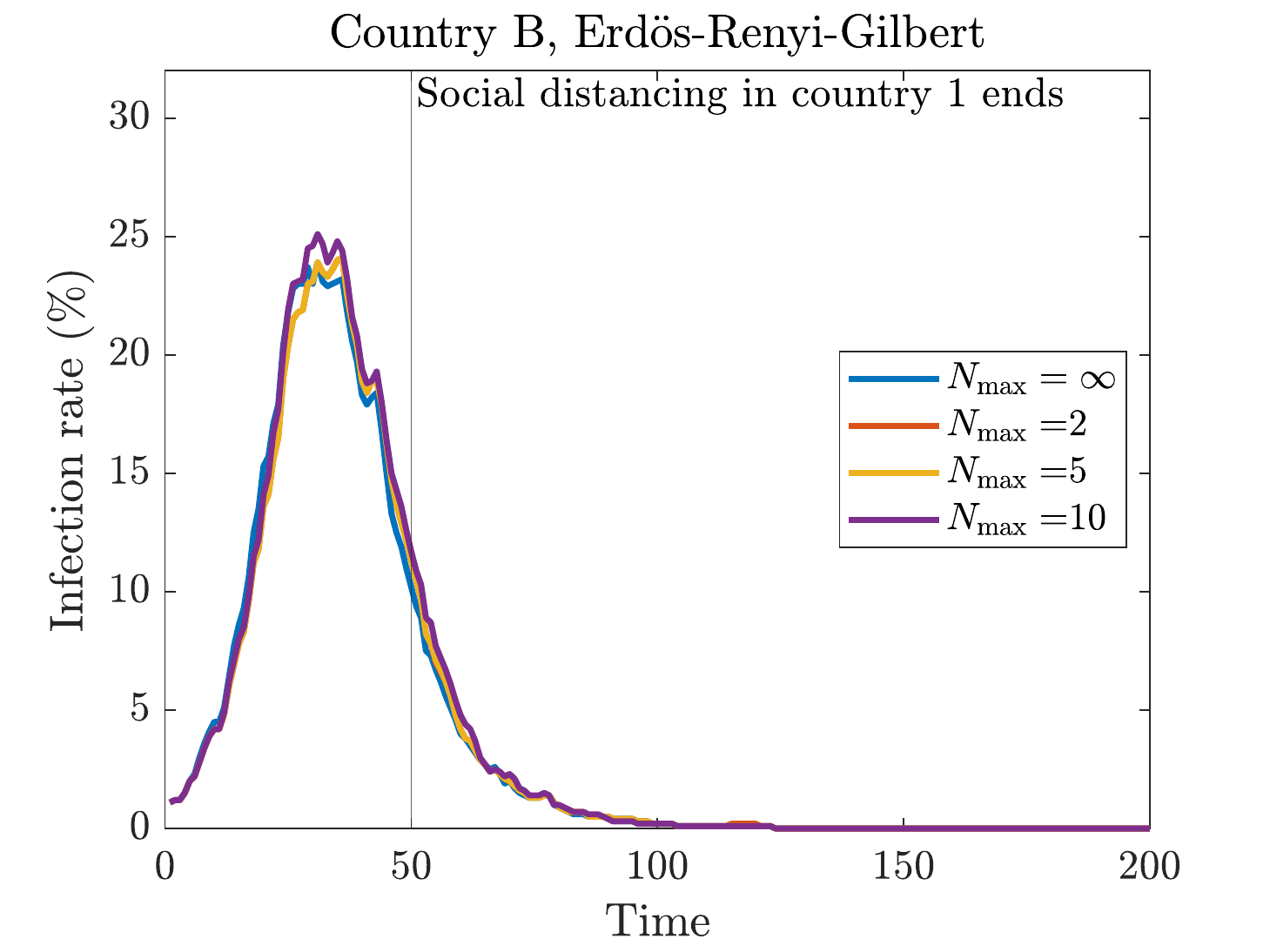}
\end{subfigure}
\begin{subfigure}{0.48\linewidth}
\includegraphics[width=\linewidth]{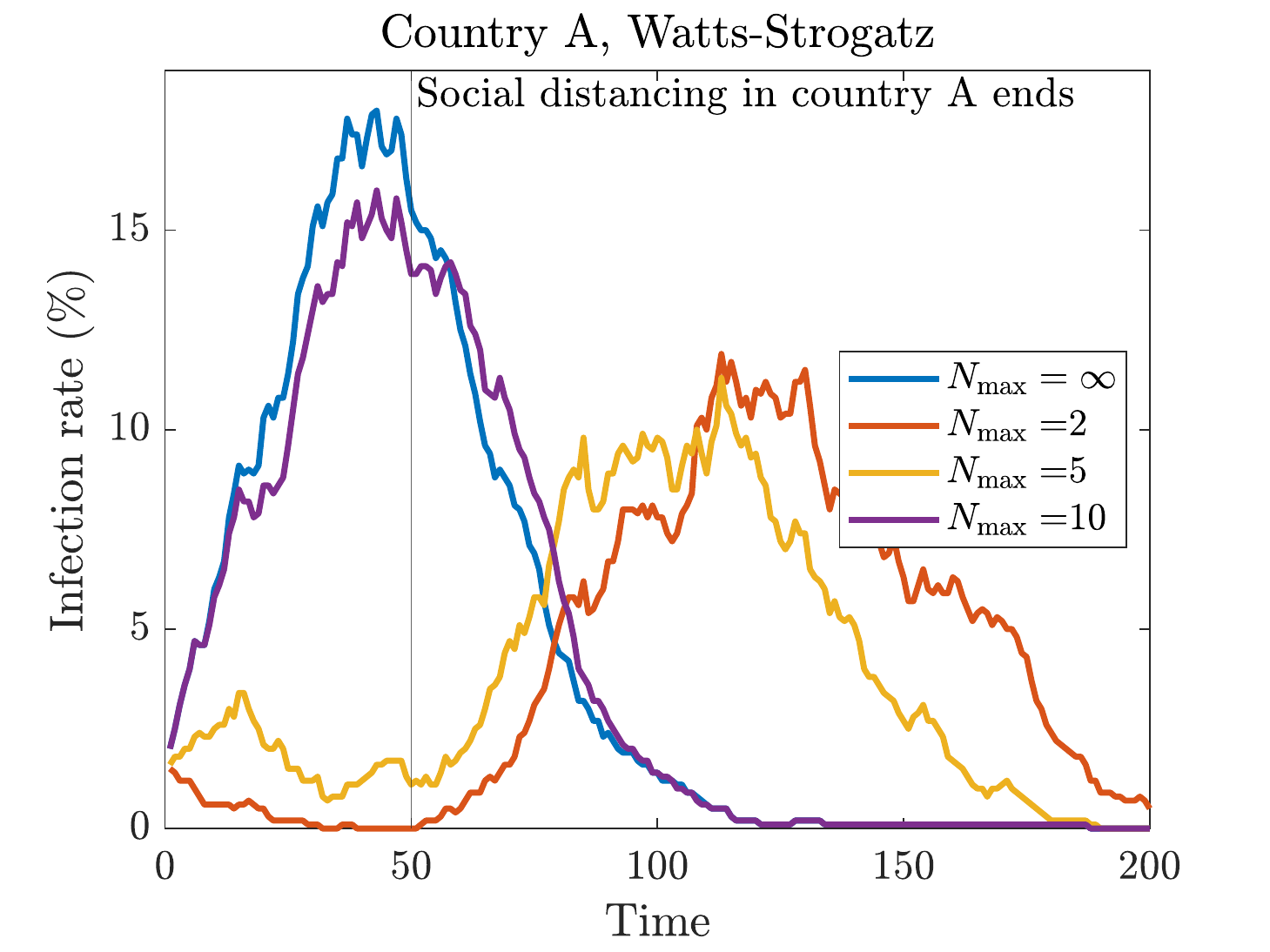}
\end{subfigure}
\begin{subfigure}{0.48\linewidth}
\includegraphics[width=\linewidth]{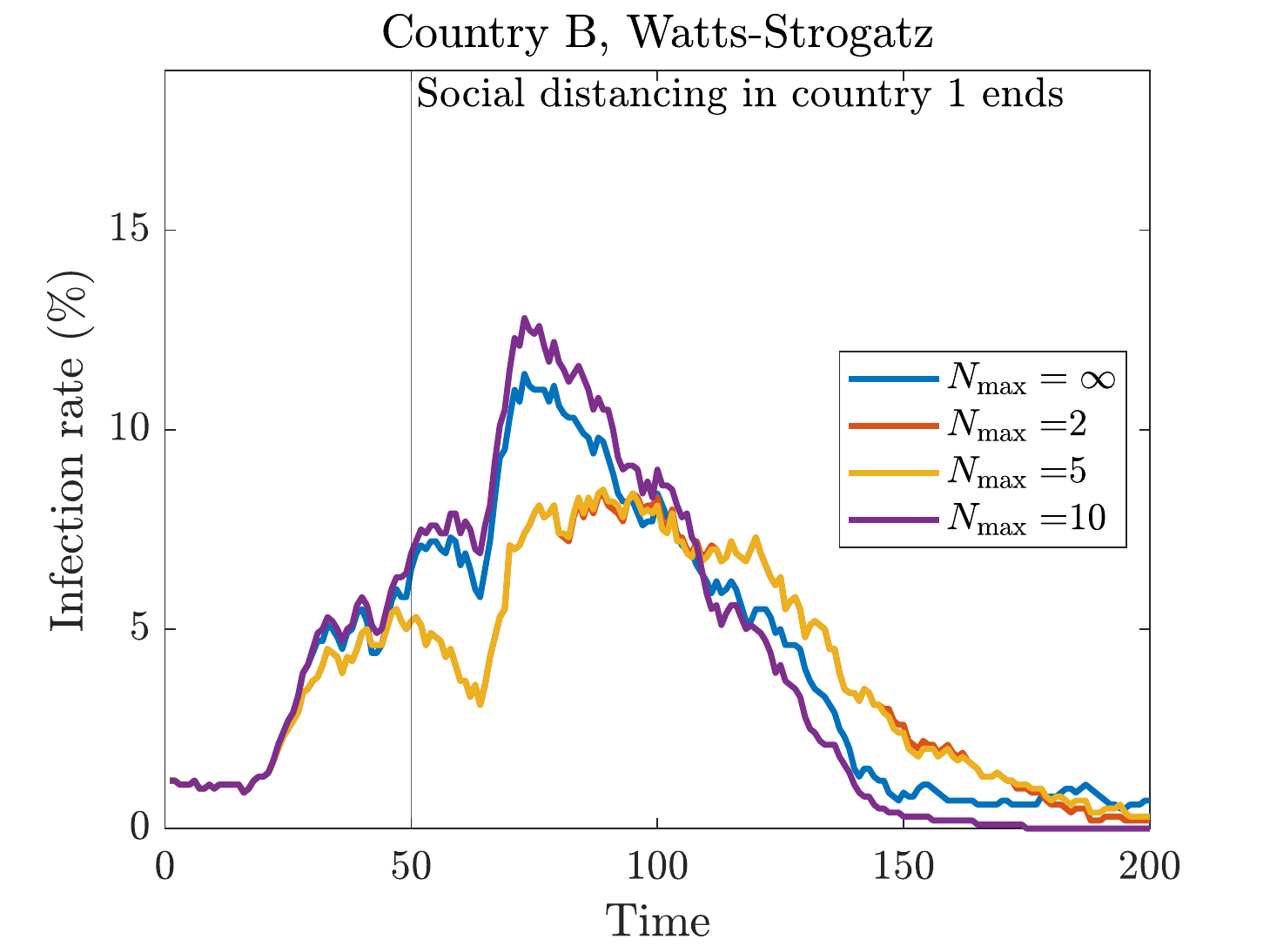}
\end{subfigure}
\begin{subfigure}{0.48\linewidth}
\includegraphics[width=\linewidth]{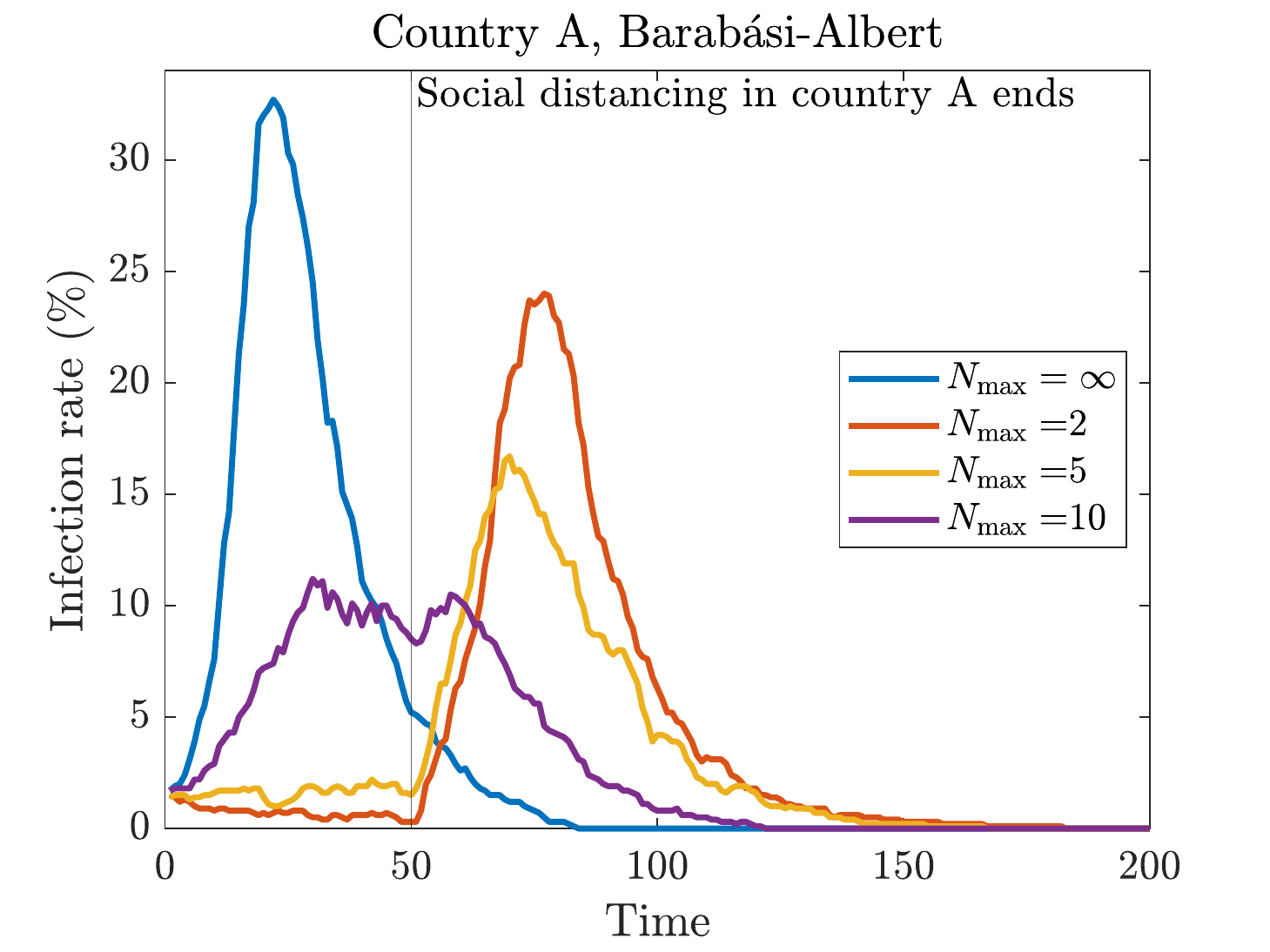}
\end{subfigure}
\begin{subfigure}{0.48\linewidth}
\includegraphics[width=\linewidth]{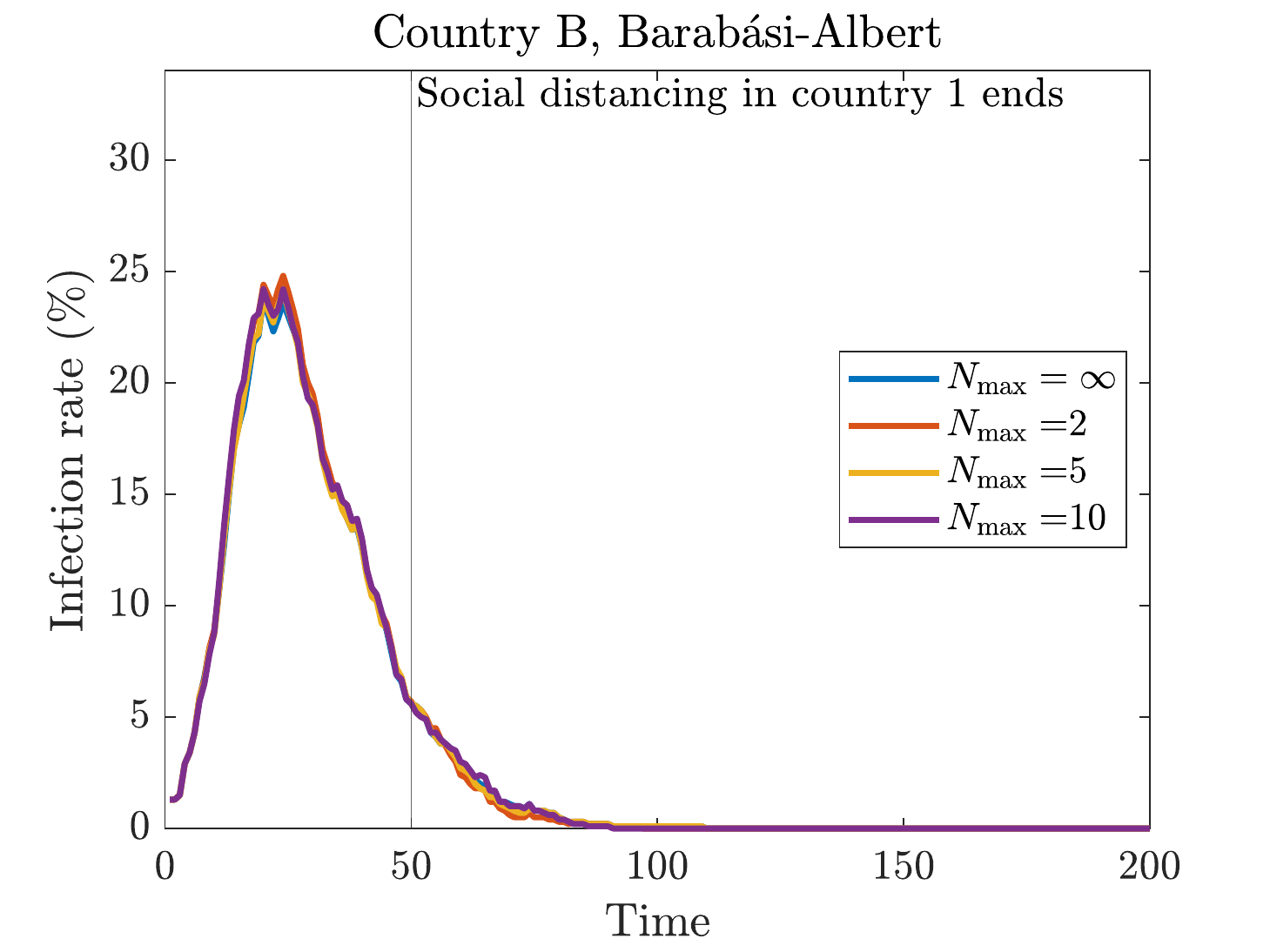}
\end{subfigure}
\caption{Epidemic dynamics with two countries.}\label{fig:twocountry}
\end{figure}

Figure \ref{fig:twocountry} contains six panels organised in a $3\times2$ format. As with Figure \ref{fig:infection_network}, the panels in a given row correspond to different network structure:\ the first refers to ERG networks, the second to WS networks, and the last row to BA networks. Within a given row, the left panel presents time-varying infection rates for Country $A$ of four separate simulations differing across social distancing measures ($\Nmax = 2, 5, 10$, and $\infty$). The vertical line at $t=50$ represents the lifting of the social distancing restriction. The right panel in each row is the corresponding time-varying infection rate for Country $B$ during the same simulation.

The interesting comparison to make is the results for Country $A$ above with that of the single country results in the left panels of Figure \ref{fig:infection_network}. Compared to the case with one country, it is clear that social distancing is less effective because new cases are imported from foreign countries. And this holds even for the strict nearly-full-lockdown social distancing measure of $\Nmax = 2$. Note that this occurs despite the fact that the expected degree for vertices in Country $A$ has increased by the seemingly negligible amount of 0.1.

These simulation results highlight the need for policies designed to tackle the COVID-19 epidemic to be coordinated at the global level. To give an extreme example, imagine a large body of interconnected individuals who live in different regions with no barriers to moving between regions (this is precisely with countries within the Schengen area of the European Union and states within the United States of America). If one region has weaker social distancing measures than all the others while maintaining connections to them, this one region can impose large negative externalities on the others, perhaps even completely wiping out the benefits one region's strict measures.


\subsection{Essential Workers}\label{subsec:Essential}

Suppose that some fraction of individuals are deemed `essential workers' who face different social distancing measures to everyone else. As a numerical example, suppose a random 10\% of the individuals in the population are essential workers, and the social distancing policy is far more lax for these individuals, specifying $\Nmax=10$ over the entire duration.

\begin{figure}[!htb]
\centering
\begin{subfigure}{0.48\linewidth}
\includegraphics[width=\linewidth]{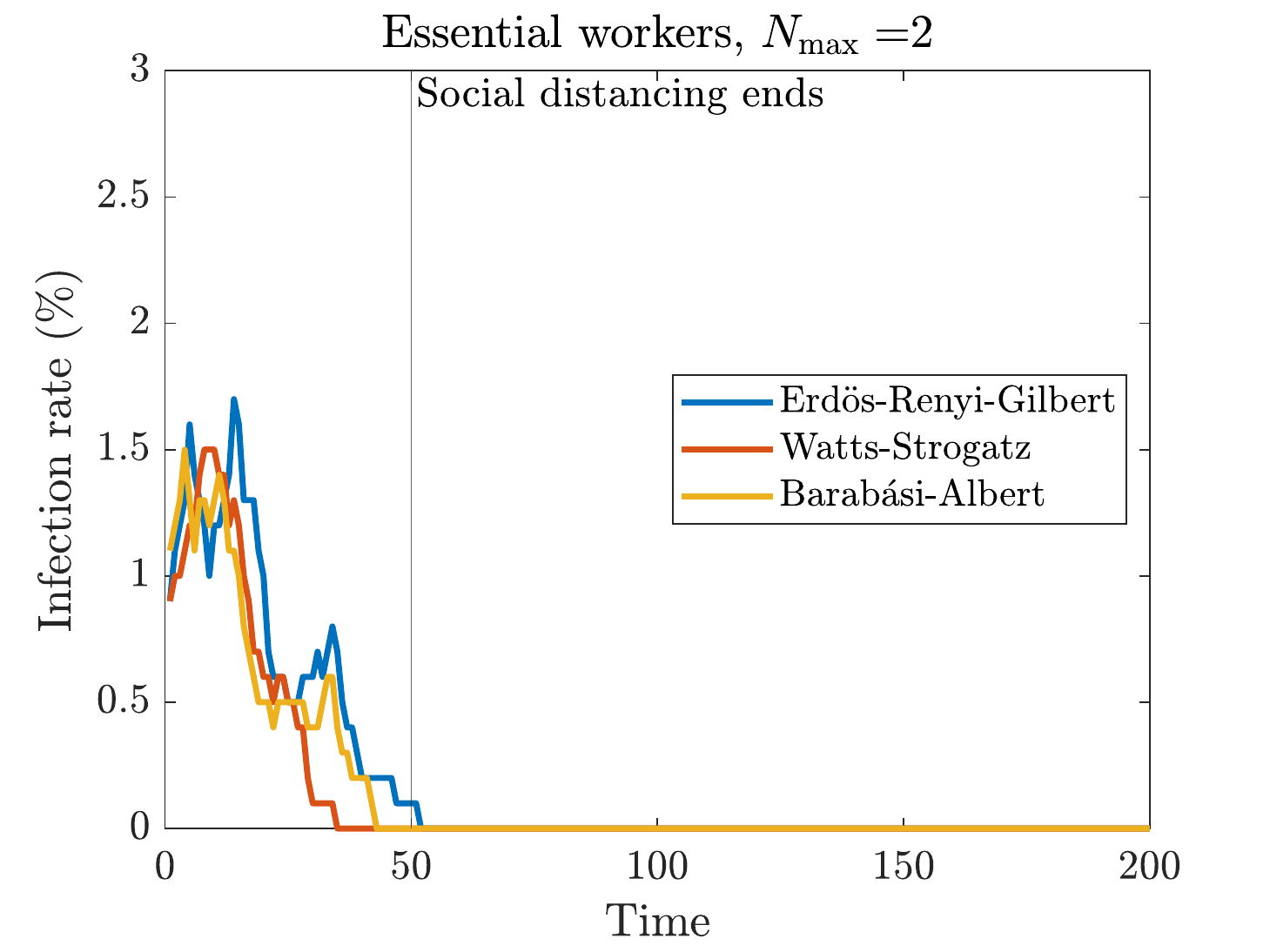}
\end{subfigure}
\begin{subfigure}{0.48\linewidth}
\includegraphics[width=\linewidth]{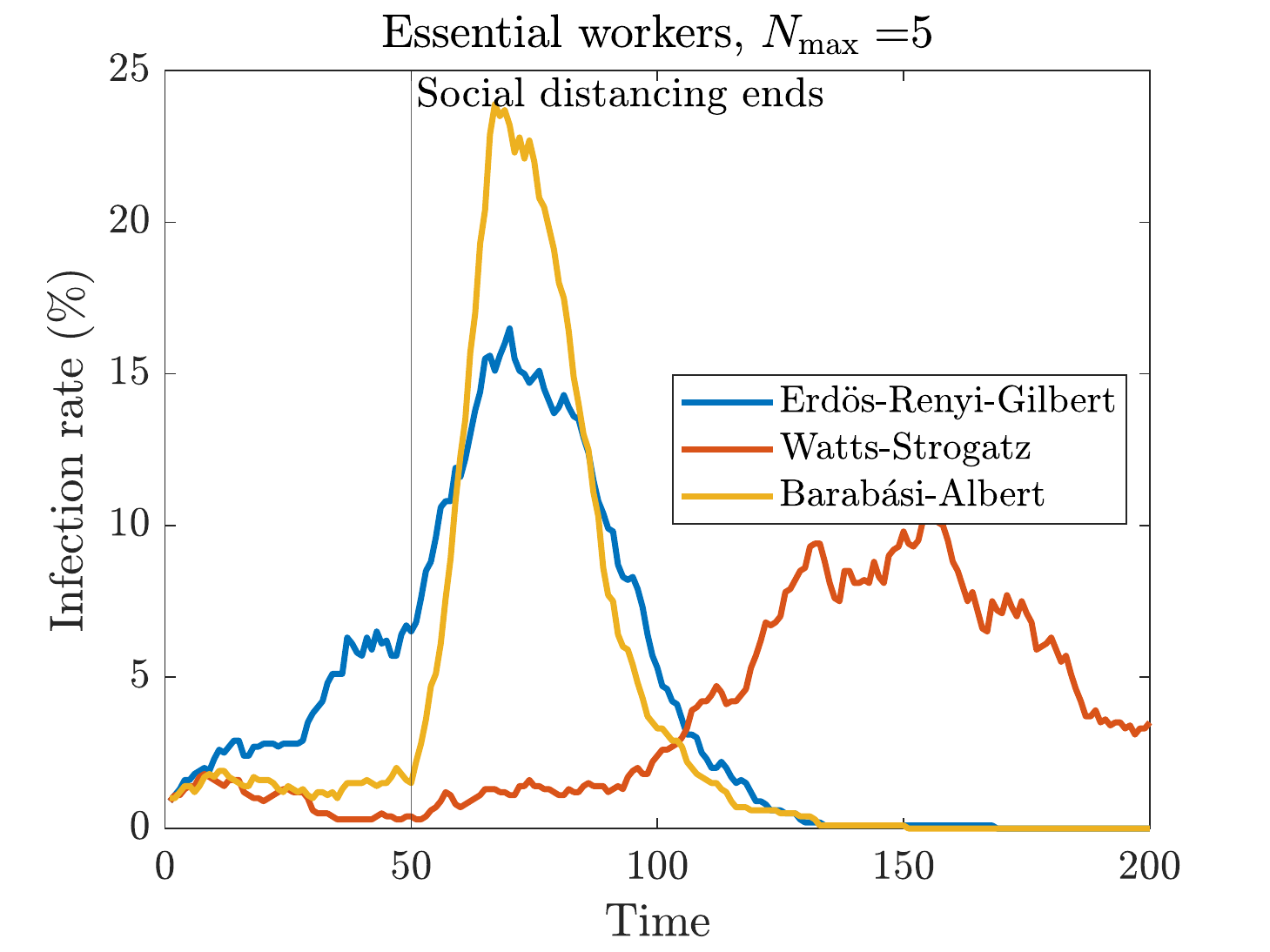}
\end{subfigure}
\caption{Epidemic dynamics with essential workers.}\label{fig:essential}
\end{figure}

Figure \ref{fig:essential} below contains two panels. In both panels 10\% of individuals are randomly designated as essential workers and assigned $\Nmax = 10$. The left panel shows the outcomes of one trial for each network type, wherein 1\% of all individuals are initially infected and the 90\% of individuals who are non-essential have $\Nmax = 2$. The right panel is similar except the 90\% of individuals who are non-essential have $\Nmax = 5$. 

In comparing the graph from each panel with the corresponding appropriate panel from Figure \ref{fig:infection_network}, it is clear that there is only a minor difference from the benchmark case where there are no essential workers. As such, if only social distancing measures were employed, they would not be very effective in curtailing the spread of the disease unless extremely strict measures are imposed on those who are deemed not essential. This seems highly relevant to the COVID-19 pandemic given that in a recent paper, \cite{McCormackAvery:2020:J} used the Public Use Microdata Sample of the 2018 American Community Survey (ACS) to estimate that one in every three jobs in the USA is deemed essential.

\subsection{Gradual Relaxation of Social Distancing}\label{subsec:Relaxing}

In tackling COVID-19, many countries introduced a ``tier structure'' of measures, where higher tiers correspond to stricter social distancing. This raises two natural policy issues:
\begin{enumerate*}
\item How many tiers should there be?
\item When should a region be permitted/forced to move down/up a tier?
\end{enumerate*}
While the second policy allows governments to respond in real time in the event of getting a bad draw, the first policy of where to place the various tiers needs to be decided in advance. Implicitly both these policies are exploiting time-varying mitigation strategies.

As an example, we consider the outcomes of two different ``tiered'' policies. They are,
\begin{description}
\item[Policy A:] Start with severe social distancing, $\Nmax = 2$, for ten days, then increment $\Nmax$ by 1 every 10 days stopping when $\Nmax = 10$. After this 90-day window, all social distancing restrictions are lifted.
\item[Policy B:] Start with mild social distancing, $\Nmax = 10$, and keep in place for 50 days. After this 50-day window, all social distancing restrictions are lifted.
\end{description}

\begin{figure}[!htb]
\centering
\includegraphics[width=0.7\linewidth]{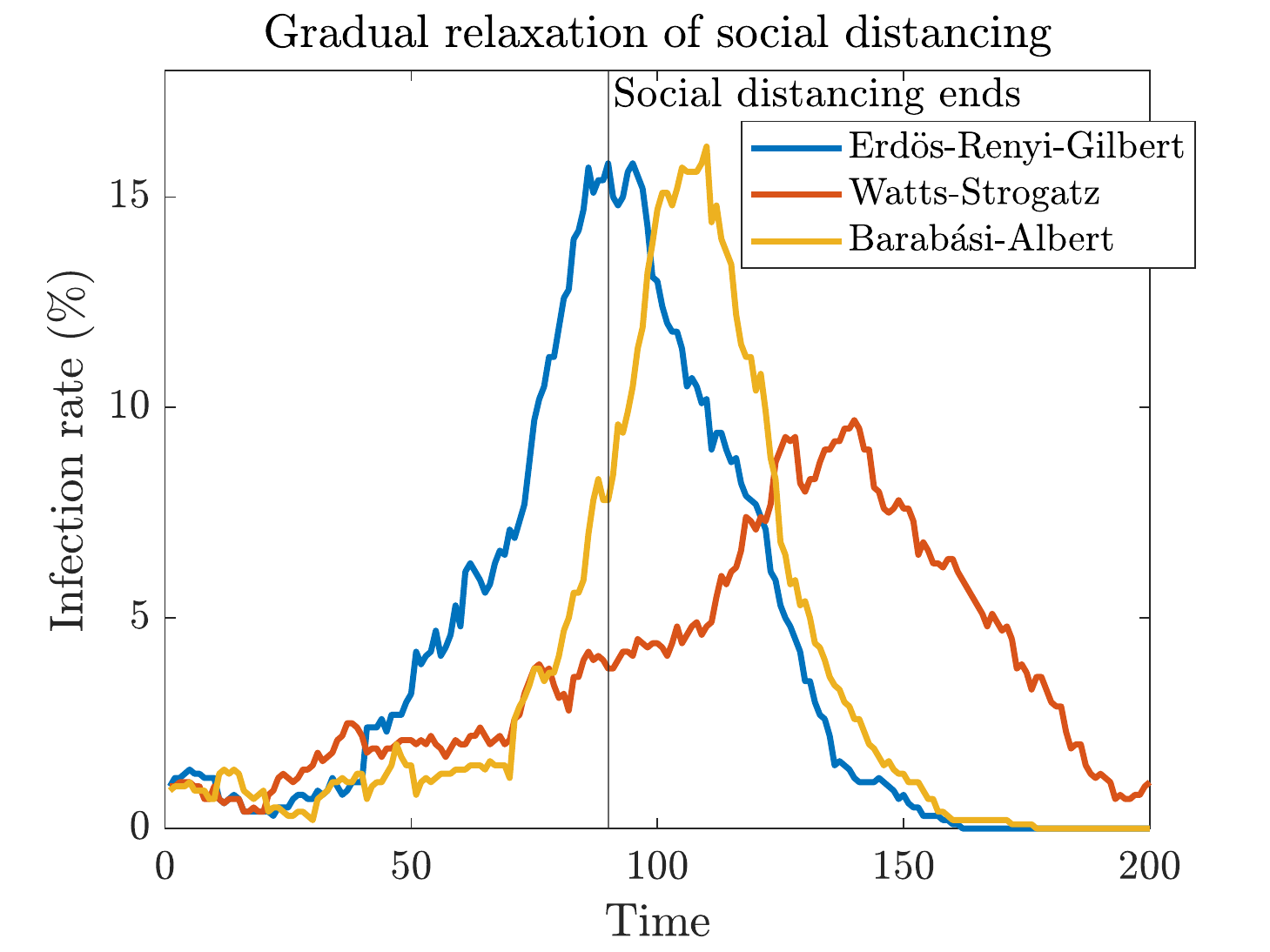}
\caption{Epidemic dynamics with gradual relaxation of social distancing.}\label{fig:gradual}
\end{figure}

Policy B above is identical to that which has already been considered (the results can be seen in Figure \ref{fig:infection_network} and Table \ref{t:large_sim}). Figure \ref{fig:gradual} presents simulation results from one trial of Policy A for each network type.

{
At least for this trial, setting $\Nmax=10$ for 50 days is considerably better for BA networks, while gradual relaxation over 90 days is better for ERG and WS. The reasons for this are similar to the results of Section \ref{sec:Results}. That is, for BA networks Policy A has the important effect of infecting the well-connected individuals but constraining their reach during their infectious period and this trumps the outcome where they are not infected until after restrictions are lifted.  
}

\section{Extensions and Related Work}\label{sec:Extensions}

Our goal with this paper is to provide a flexible framework that allows policy makers to simulate precisely how social distancing can mitigate the spread of an infectious pathogen. The model we have proposed is the standard SIR model on a network with social distancing constraints imposed upon individuals, that we call the SIRwSD model. While we have considered some such social distancing policies in Section \ref{sec:PolicyQs} above, there are of course many more. We briefly sketch some of these now.

\paragraph{Other social distancing models.}
Perhaps the paper closest to ours is \cite{MaharajKleczkowski:2012:PH}. These authors consider a networked SIR model on a square lattice. They introduce two radii:\ the first radius defines an individual's neighbourhood and the second radius defines the individual's neighbourhood with social distancing measures in effect. The main differences with our model and this is that we allow individual-specific social distancing restrictions, and we do not limit analysis to networks where all individuals are in some sense `the same'.

\paragraph{Random interactions.}
Even during a full lockdown, setting $\kappa(i) = 0$ for every individual $i$, people are still permitted to make outings for essential items. As such, there is always the possibility that an individual will contract a disease from someone with whom they are not social acquaintances. This feature could easily be included in the model by allowing the possibility of interactions that are not specified by the contact network.



\paragraph{Asymmetries in transmission probabilities.}
We have assumed that the transmission probability $\beta$ is the same for all individuals. But there is medical literature that suggests that this is not so for COVID-19 \citep{IypeGulati:2020:M, ZhouLi:2020:E}. We have further assumed that the recovery probability $\gamma$ is constant for everyone. But, at least as regards deaths attributed to the COVID-19 pathogen, this is not accurate, with the likelihood of recovery decreasing with age.\footnote{It is further possible that an individual's likelihood of recovery may be decreasing in the number of those currently infected. The reason being that most countries have a fixed stock of medical resources, so additional hospitalisations reduce the per patient allocation of resources, which once a threshold is met will have ramifications for quality of treatment.} Both of these richer features can easily be incorporated into our setup.

\paragraph{Directed connections.}
Most models of how epidemics spread in social networks assume the underlying graph is undirected. This implies that the individuals most likely to spread the disease are also the most likely to become infected. \cite{AllardMoore:2020:} relax this assumption allowing the underlying graph to be directed (and not just a complete biorientation of an undirected graph, so that the in-degree of a vertex need not equal its out-degree). \cite{AllardMoore:2020:} show that this can have policy implications for both contact tracing and the prevention of superspreading events. Our social distancing measure can easily be appended to a ``directed'' framework by constraining each vertex $v$ according to the arcs with tail at $v$ (the out-edges of vertex $v$).

\paragraph{Reinfection is possible.}
The $R$ state in the SIR model refers to ``Removed'', but is more accurately described as ``Removed and no longer Susceptible''. As regards COVID-19, there is new data emerging from South Korea that indicates that some who have been infected may be prone to reinfection. If so, the spread of the coronavirus may be better described by an SIS model (where the second `S' also stands for susceptible). This important change is easily incorporated to our setup.

\paragraph{Asymmetries in global travel and `Travel Corridors'.}
Connections between counties need not always be equally distributed as we assumed in Subsection \ref{subsec:Global}. For example, the number of UK citizens who holiday in Spain dwarfs the number of Spanish citizens who holiday in the UK. Such a setup could be incorporated into our framework by assuming that a large number of UK citizens (the holiday makers) each have a few connection to a small number of Spanish citizens (those who work in hospitality). How such asymmetries in connections play out may be important in deciding which `travel corridors' are safer than others.\footnote{In the European Union, it is not uncommon to see the following occur:\ if the state of the pandemic in Country A is deemed sufficiently safe by Country B, then Country B will allow a `travel corridor' from Country A. In practice, a travel corridor from A to B means that Country B will allow visitors who have been to country A in the last 14 days; in our framework it would mean allowing (possibly directed) connections from A to B.} Such policies are easily explored using our framework.

\paragraph{Individual specific policies:\ key to target the right individuals.}
As mentioned before, a contact network $G$ and social distancing policy $\kappa$ together generate an infectious subnetwork. Since the social distancing policy $\kappa$ temporarily deletes a subset of edges from $G$, there are simply less avenues along which the disease may be transmitted. However, perhaps counterintuitively this will not always reduce the reach of an epidemic. While our results in Section \ref{sec:Results} show that the reach is reduced when all individuals in society face the same constraints, \cite{MuscilloPin:2020:A} show that if those individuals who have many neighbours reduce their contacts proportionately less than those who have few, then the disease can take longer to die out. The specification considered by \cite{MuscilloPin:2020:A} is a special case of our framework where $\kappa$ is chosen such that the fraction $\kappa(i)/d(i)$ is increasing in degree.


\paragraph{Interplay between time-varying and individual-specific constraints.}
In Subsections \ref{subsec:Global} and \ref{subsec:Essential}, we allowed social distancing constraints to be group-specific (i.e., varying from individual to individual). In Subsection \ref{subsec:Relaxing} we considered time-varying policies. How group-specific constraints and time-varying constraints interact together is not something we have explored in this paper but can easily be handled within our framework.

\paragraph{Full compliance and rational individuals.}
In our model, a social distancing policy is set and all individuals adhere to it. Clearly this a strong assumption as full compliance cannot be guaranteed without strong monitoring. We can easily incorporate this feature in our model is by defining a parameter $\xi \in (0, 1)$ where $\xi$ is the probability that an individual violates the restrictions in a given period. A related point is made by the game-theoretic literature on epidemic control. In a game-theoretic model, individuals are assumed to be rational and rational individuals may themselves take mitigating measures. In such a setup, increased social distancing reduces likelihood of infection but also comes with a cost in the form of reduced freedom, convenience, earning potential and so on.  \cite{Reluga:2010:PCB} considers a differential game capturing precisely these trade-offs. \cite{ValdezMacri:2012:PRSNSMP} allow for the possibility that individuals can identify infected neighbours and temporarily interrupt contact with them.

%


\paragraph{Incorporating social distancing policies into an economic model.}
Lastly, while we have focused on the benefits to public health of social distancing, this is far from the full story. Any amount of social distancing brings with it economic cost and, as evidenced by recent economic data, the policy responses to the COVID-19 outbreak have had an enormous effect on the global economy.\footnote{\cite{BakerBloom:2020:} document that the USA's week-ending jobless claims numbers from late-March 2020 have been an order of magnitude higher than any seen previously. Moreover, these numbers were as a result of the lockdown policy and not the virus itself.} Given that, as of time of writing, the mass roll-out of a successful vaccine to COVID-19 is potentially far on the horizon, some form of relaxed social distancing measures need to be considered.\footnote{\cite{DingelNeiman:2020:CEVRP} estimate that only one-third of jobs in the USA can be done from home.} A richer model would also include the effect that social distancing has on economic output, and subsequently incorporate the interdependent relationship between output and public health \citep{BloomCanning:2004:WD,Weil:2007:QJE}. Such a model would allow policy makers to consider the full trade off between absolute lockdown (everyone safe from exposure, economic activity greatly reduced) versus no lockdown of any kind (nobody safe from exposure, economic activity is maximal).

\paragraph{}
Lastly, we wish to reiterate that the framework in this paper considers the benefits to social distancing when it is the only policy available tool available.  Clearly this is not realistic. In real world epidemics policy makers have an array of tools available. While understanding how one policy works in isolation is a useful theoretical exercise, it is important to understand how the full set of policy tools complement each other in order to successfully operationalise a pandemic response plan.

\vspace{.1in}

\noindent
Gregory Gutin has nothing to disclose. Sung-Ha Hwang has nothing to disclose. Tomohiro Hirano has nothing to disclose. Philip R Neary has nothing to disclose. Alexis Akira Toda has nothing to disclose.

%
%
%
%
%
%
%

\newpage
\bibliographystyle{plainnat}
\bibliography{covidLibrary}

\begin{thebibliography}{26}
\providecommand{\natexlab}[1]{#1}
\providecommand{\url}[1]{\texttt{#1}}
\expandafter\ifx\csname urlstyle\endcsname\relax
  \providecommand{\doi}[1]{doi: #1}\else
  \providecommand{\doi}{doi: \begingroup \urlstyle{rm}\Url}\fi

\bibitem[Allard et~al.(2020)Allard, Moore, Scarpino, Althouse, and
  H{\'e}bert-Dufresne]{AllardMoore:2020:}
Antoine Allard, Cristopher Moore, Samuel~V. Scarpino, Benjamin~M. Althouse, and
  Laurent H{\'e}bert-Dufresne.
\newblock The role of directionality, heterogeneity and correlations in
  epidemic risk and spread, 2020.
\newblock URL \url{https://arxiv.org/abs/2005.11283}.

\bibitem[Baker et~al.(2020)Baker, Bloom, Davis, Kost, Sammon, and
  Tasaneeya]{BakerBloom:2020:}
Scott~R. Baker, Nicholas Bloom, Steven~J. Davis, Kyle Kost, Marco Sammon, and
  Viratyosin Tasaneeya.
\newblock The unprecedented stock market impact of {COVID}-19.
\newblock \emph{Covid Economics, Vetted and Real-Time Papers}, 1, 2020.

\bibitem[Barab{\'a}si and Albert(1999)]{BarabasiAlbert:1999:S}
Albert-L{\'a}szl{\'o} Barab{\'a}si and R{\'e}ka Albert.
\newblock Emergence of scaling in random networks.
\newblock \emph{Science}, 286\penalty0 (5439):\penalty0 509, October 1999.
\newblock \doi{10.1126/science.286.5439.509}.

\bibitem[Bloom et~al.(2004)Bloom, Canning, and Sevilla]{BloomCanning:2004:WD}
David~E. Bloom, David Canning, and Jaypee Sevilla.
\newblock The effect of health on economic growth: A production function
  approach.
\newblock \emph{World Development}, 32\penalty0 (1):\penalty0 1 -- 13, January
  2004.
\newblock \doi{10.1016/j.worlddev.2003.07.002}.

\bibitem[Dingel and Neiman(2020)]{DingelNeiman:2020:CEVRP}
Jonathan Dingel and Brent Neiman.
\newblock How many jobs can be done at home?
\newblock \emph{Journal of Public Economics}, 189:\penalty0 104235, September
  2020.
\newblock \doi{10.1016/j.jpubeco.2020.104235}.

\bibitem[Erd\H{o}s and R\'enyi(1959)]{ErdosRenyi:1959:PMD}
Paul Erd\H{o}s and Alfr\'ed R\'enyi.
\newblock On random graphs {I}.
\newblock \emph{Publicationes Mathematicae Debrecen}, 6:\penalty0 290--297,
  1959.

\bibitem[Gerke et~al.(2019)Gerke, Gutin, Hwang, and Neary]{GerkeGutin:2019:A}
Stefanie Gerke, Gregory Gutin, Sung-Ha Hwang, and Philip~R. Neary.
\newblock {N}etflix games: Local public goods with capacity constraints.
\newblock 2019.
\newblock URL \url{https://arxiv.org/abs/1905.01693}.

\bibitem[Gilbert(1959)]{Gilbert:1959:AMS}
Edgar~N. Gilbert.
\newblock Random graphs.
\newblock \emph{Annals of Mathematical Statistics}, 30\penalty0 (4):\penalty0
  1141--1144, 1959.
\newblock \doi{10.1214/aoms/1177706098}.

\bibitem[Gutin et~al.(2020)Gutin, Neary, and Yeo]{GutinNeary:2020:A}
Gregory Gutin, Philip~R. Neary, and Anders Yeo.
\newblock Uniqueness of {$DP$}-{N}ash subgraphs and {$D$}-sets in weighted
  graphs of {N}etflix games.
\newblock In Donghyun Kim, R.~N. Uma, Zhipeng Cai, and Dong~Hoon Lee, editors,
  \emph{Computing and Combinatorics}, volume 12273 of \emph{Lecture Notes in
  Computer Science}, pages 360--371. Springer, 2020.
\newblock \doi{10.1007/978-3-030-58150-3_29}.

\bibitem[Iype and Gulati(2020)]{IypeGulati:2020:M}
Eldhose Iype and Sadhya Gulati.
\newblock Understanding the asymmetric spread and case fatality rate ({CFR})
  for {COVID}-19 among countries.
\newblock 2020.
\newblock URL
  \url{https://www.medrxiv.org/content/early/2020/04/26/2020.04.21.20073791}.

\bibitem[Kermack and Mc{K}endrick(1927)]{KermackMcKendrick:1927:PRSLSCPMPC}
William~O. Kermack and Anderson~G. Mc{K}endrick.
\newblock A contribution to the mathematical theory of epidemics.
\newblock \emph{Proceedings of the Royal Society A}, 115\penalty0
  (772):\penalty0 700--721, August 1927.
\newblock \doi{10.1098/rspa.1927.0118}.

\bibitem[Kiss et~al.(2017)Kiss, Miller, and Simon]{KissMiller:2017:}
Istv{\'a}n~Z. Kiss, Joel~C. Miller, and P{\'e}ter~L. Simon.
\newblock \emph{Mathematics of Epidemics on Networks}.
\newblock Springer, 2017.

\bibitem[Lan{\v{c}}i{\'{c}} et~al.(2011)Lan{\v{c}}i{\'{c}}, Antulov-Fantulin,
  {\v{S}}iki{\'{c}}, and
  {\v{S}}tefan{\v{c}}i{\'{c}}]{LancicAntulov-Fantulin:2011:PSMA}
Alen Lan{\v{c}}i{\'{c}}, Nino Antulov-Fantulin, Mile {\v{S}}iki{\'{c}}, and
  Hrvoje {\v{S}}tefan{\v{c}}i{\'{c}}.
\newblock Phase diagram of epidemic spreading---unimodal vs. bimodal
  probability distributions.
\newblock \emph{Physica A: Statistical Mechanics and its Applications},
  390\penalty0 (1):\penalty0 65--76, January 2011.
\newblock \doi{10.1016/j.physa.2010.06.024}.

\bibitem[Li et~al.(2020)Li, Guan, Wu, Wang, Zhou, Tong, Ren, Leung, Lau, Wong,
  Xing, Xiang, Wu, Li, Chen, Li, Liu, Zhao, Liu, Tu, Chen, Jin, Yang, Wang,
  Zhou, Wang, Liu, Luo, Liu, Shao, Li, Tao, Yang, Deng, Liu, Ma, Zhang, Shi,
  Lam, Wu, Gao, Cowling, Yang, Leung, and Feng]{Li_2020}
Qun Li, Xuhua Guan, Peng Wu, Xiaoye Wang, Lei Zhou, Yeqing Tong, Ruiqi Ren,
  Kathy~S.M. Leung, Eric~H.Y. Lau, Jessica~Y. Wong, Xuesen Xing, Nijuan Xiang,
  Yang Wu, Chao Li, Qi~Chen, Dan Li, Tian Liu, Jing Zhao, Man Liu, Wenxiao Tu,
  Chuding Chen, Lianmei Jin, Rui Yang, Qi~Wang, Suhua Zhou, Rui Wang, Hui Liu,
  Yinbo Luo, Yuan Liu, Ge~Shao, Huan Li, Zhongfa Tao, Yang Yang, Zhiqiang Deng,
  Boxi Liu, Zhitao Ma, Yanping Zhang, Guoqing Shi, Tommy~T.Y. Lam, Joseph~T.
  Wu, George~F. Gao, Benjamin~J. Cowling, Bo~Yang, Gabriel~M. Leung, and Zijian
  Feng.
\newblock Early transmission dynamics in {W}uhan, {C}hina, of novel
  coronavirus-infected pneumonia.
\newblock \emph{New England Journal of Medicine}, 382:\penalty0 1199--1207,
  March 2020.
\newblock \doi{10.1056/NEJMoa2001316}.

\bibitem[Maharaj and Kleczkowski(2012)]{MaharajKleczkowski:2012:PH}
Savi Maharaj and Adam Kleczkowski.
\newblock Controlling epidemic spread by social distancing: Do it well or not
  at all.
\newblock \emph{{BMC} Public Health}, 12\penalty0 (1):\penalty0 679, August
  2012.
\newblock \doi{10.1186/1471-2458-12-679}.

\bibitem[Mc{C}ormack et~al.(2020)Mc{C}ormack, Avery, Kahn-Lang~Spitzer, and
  Chandra]{McCormackAvery:2020:J}
Grace Mc{C}ormack, Christopher Avery, Ariella Kahn-Lang~Spitzer, and Amitabh
  Chandra.
\newblock Economic vulnerability of households with essential workers.
\newblock \emph{{JAMA}}, 324\penalty0 (4):\penalty0 388--390, 2020.
\newblock \doi{10.1001/jama.2020.11366}.

\bibitem[Moreno et~al.(2002)Moreno, Pastor-Satorras, and
  Vespignani]{MorenoPastor-Satorras:2002:EPJCMCS}
Y.~Moreno, R.~Pastor-Satorras, and A.~Vespignani.
\newblock Epidemic outbreaks in complex heterogeneous networks.
\newblock \emph{The European Physical Journal B - Condensed Matter and Complex
  Systems}, 26\penalty0 (4):\penalty0 521--529, 2002.
\newblock \doi{10.1140/epjb/e20020122}.
\newblock URL \url{https://doi.org/10.1140/epjb/e20020122}.

\bibitem[Muscillo et~al.(2020)Muscillo, Pin, and Razzolini]{MuscilloPin:2020:A}
Alessio Muscillo, Paolo Pin, and Tiziano Razzolini.
\newblock {C}ovid19: Unless one gets everyone to act, policies may be
  ineffective or even backfire.
\newblock \emph{{PLOS} {ONE}}, 15\penalty0 (9):\penalty0 e0237057, 2020.
\newblock \doi{10.1371/journal.pone.0237057}.

\bibitem[Pastor-Satorras and
  Vespignani(2001)]{Pastor-SatorrasVespignani:2001:PRL}
Romualdo Pastor-Satorras and Alessandro Vespignani.
\newblock Epidemic spreading in scale-free networks.
\newblock \emph{Phys. Rev. Lett.}, 86:\penalty0 3200--3203, Apr 2001.
\newblock \doi{10.1103/PhysRevLett.86.3200}.
\newblock URL \url{https://link.aps.org/doi/10.1103/PhysRevLett.86.3200}.

\bibitem[Reluga(2010)]{Reluga:2010:PCB}
Timothy~C. Reluga.
\newblock Game theory of social distancing in response to an epidemic.
\newblock \emph{{PLoS} Computational Biology}, 6\penalty0 (5):\penalty0
  e1000793, May 2010.
\newblock \doi{10.1371/journal.pcbi.1000793}.

\bibitem[Toda(2020)]{Toda:2020:}
Alexis~Akira Toda.
\newblock Susceptible-infected-recovered ({SIR}) dynamics of {COVID}-19 and
  economic impact.
\newblock Covid Economics~1, CEPR, 2020.
\newblock URL \url{https://arxiv.org/abs/2003.11221}.

\bibitem[Valdez et~al.(2012)Valdez, Macri, and
  Braunstein]{ValdezMacri:2012:PRSNSMP}
L.~Valdez, Pablo Macri, and Lidia Braunstein.
\newblock Intermittent social distancing strategy for epidemic control.
\newblock \emph{Physical Review E}, 85\penalty0 (3):\penalty0 036108, 2012.
\newblock \doi{10.1103/PhysRevE.85.036108}.

\bibitem[Watts and Strogatz(1998)]{WattsStrogatz:1998:N}
Duncan~J. Watts and Steven~H. Strogatz.
\newblock Collective dynamics of `small-world' networks.
\newblock \emph{Nature}, 393\penalty0 (6684):\penalty0 440--442, June 1998.
\newblock \doi{10.1038/30918}.

\bibitem[Weil(2007)]{Weil:2007:QJE}
David~N. Weil.
\newblock Accounting for the effect of health on economic growth.
\newblock \emph{Quarterly Journal of Economics}, 122\penalty0 (3):\penalty0
  1265--1306, August 2007.
\newblock \doi{10.1162/qjec.122.3.1265}.

\bibitem[Zhaoyang et~al.(2018)Zhaoyang, Sliwinski, Martire, and
  Smyth]{ZhaoyangSliwinski:2018:PA}
Ruixue Zhaoyang, Martin Sliwinski, Lynn Martire, and Joshua Smyth.
\newblock Age differences in adults daily social interactions: An ecological
  momentary assessment study.
\newblock \emph{Psychology and Aging}, 33\penalty0 (4):\penalty0 607--618,
  2018.
\newblock \doi{10.1037/pag0000242}.

\bibitem[Zhou et~al.(2020)Zhou, Li, Lu, Ma, Liu, Zhu, Hu, Wu, Chen, Wang, Wei,
  Li, Xu, and Wang]{ZhouLi:2020:E}
Fuling Zhou, Jingfeng Li, Mengxin Lu, Pan~Yunbao Ma, Linlu~and, Xiaoyan Liu,
  Xiaobin Zhu, Chao Hu, Sanyun Wu, Liangjun Chen, Yi~Wang, Yongchang Wei,
  Yirong Li, Haibo Xu, and Cai~Lin Wang, Xinghuan~and.
\newblock Tracing asymptomatic {SARS-CoV-2} carriers among 3674 hospital staff:
  A cross-sectional survey.
\newblock \emph{EClinicalMedicine}, 26:\penalty0 100510, September 2020.
\newblock \doi{10.1016/j.eclinm.2020.100510}.

\end{thebibliography}

\end{document}